\documentclass[preprint,aps,prc,eqsecnum,
showpacs,showkeys,draft] {revtex4-1}
\def\bfgr #1{ \mbox {{\boldmath $#1$}}}
\usepackage{axodraw}
\usepackage{lineno}
\usepackage{Citesort}
\usepackage{bm}
\linenumbers[1]
\begin{document}
\setcounter{page}{0}
\title{On the usage of electron beam as a tool to produce
radioactive isotopes in photo-nuclear reactions.}
\author{G.  G.  Bunatian \footnote{Email: bunat@cv.jinr.dubna.ru},
V. G. Nikolenko and A. B. Popov}
\affiliation{\bf Joint Institute for Nuclear Research,
 141980, Dubna, Russia}
\date{\today}
\begin{abstract} We treat the bremsstrahlung induced by initial
electron beam in converter, and the production of a desirable
radio-isotope due to the photo-nuclear reaction caused by this
bremsstrahlung. By way of illustration, the yield of a number of
some, the most in practice applicable, radio-isotopes is evaluated.
The acquired findings persuade us that usage of modern electron
accelerators offers a practicable way to produce the radio-isotopes
needful nowadays for various valuable applications in the nuclear
medicine.
\end{abstract}
\maketitle
\setcounter{page}{1}
\setcounter{section}{1}
\section*{1. I\lowercase{ntroduction}}
\label{sec:level1}
\setcounter{equation}{0}
\begin{linenumbers}
 Nowadays, it is rather impossible to find any branch of science,
industry, medicine, forensic, asf, in which the radio-isotopes are
not widely used \cite{int3,int4,int5,int7,ind,isi}. Although there
is a series of unstable natural isotopes arising from the decay of
primordial uranium and thorium, the most of about 200
radio-isotopes used for now on a regular basis are produced
artificially. Despite all the nuclear reactors produce the manifold
radio-isotopes as a result of fission of $^{235}\mbox{U}$ contained
in their fuels, the recovery of these radio-isotopes is extremely
problematic issue, and they would not be received for primary
applications, especially for medical use. At present radio-isotope
marketable production is primarily brought about by exposition of
the appropriate element to neutrons in a nuclear reactor, or to
charged particles, like protons, deuterons, or alpha particles, in
a cyclotron \cite{hu}. As a general rule, it is far more difficult
to make a radio-isotope in a cyclotron than in a reactor. Cyclotron
nuclear reactions are less productive and less predictable than
ones performed in a reactor. The variety of cyclotron-produced
radio-isotopes is tightly restricted too. Economic factors would
also militate against cyclotron production. In fact, it proves to
be anyway not competitive with the reactor radio-isotope
production.

As to reactor-based manufacturing, there are two processes to
produce isotopes: fission of $^{235}\mbox{U}$ by neutrons within an
exposed target, with subsequent recovery of a desirable isotope out
of fission fragments, and neutron capture by nucleus of an
appropriate sample, which results in elaboration of a required
isotope \cite{hu}. The $^{235}\mbox{U}$ fission cross-section
$\sigma_{nf}$ is well known to be at least a factor of about 100
greater than the typical neutron capture
$A(Z,N)(n,\gamma)A'(Z,N+1)$ cross-section to produce some
radio-isotope $A'(Z,N+1)$ that could other-ways be recovered from
$^{235}\mbox{U}$ fission fragments. That is why the radio-isotope
consumers community world-through dismissed the neutron capture as
a viable process for production of the primary needful
radio-isotopes in quantities required to meet global demand,
thought this process could be used to make minor radio-isotope
amounts to provide a stable domestic supply. For instance, in
Russia different radio-isotopes, including
$^{99}\mbox{Mo}{/}^{99m}\mbox{Tc}$, are produced on the Leningrad
power station using the neutron capture reactions in the channel of
the $RBMK$-1000 reactor \cite{is1}.

Thus, in these days, the most of world's production of primary
radio-isotopes is carried out by irradiating highly enriched
uranium ($HEU$) targets in research and test reactors that provide
a thermal neutron flux $(10^{14}-10^{15}){\mbox{n}}{/}{(\mbox{sec
cm}^2)}$, and are fueled with low enriched uranium ($LEU$), or in
some cases with ($HEU$) as well \cite{hu}. These reactors have
become indispensable for the industrial production of marketable
radio-isotopes, in particular medical isotopes to supply the
rapidly increasing demand for diagnostic and therapeutic procedures
based on nuclear medicine techniques. The nuclear-medicine
community defines the medical isotopes to include first of all the
isotope $^{99}\mbox{Mo}$, that is the precursor to the short-living
$^{99m}\mbox{Tc}$ that's used in $\approx 85\%$ of all the nuclear
medicine procedures worldwide, and also $^{131}\mbox{I} ,
^{133}\mbox{X}$ and other manifold radioactive materials
used to produced radiopharmaceuticals
\cite{int4,int7,hu,int2,int1}. The medical radio-isotope recovery
is humanly the most vital outcome of nuclear physics and industry.

The supply reliability of radio-pharmacies, hospitals, clinics and
outpatients centers with the radio-isotopes is currently the
primary concern of world nuclear-medicine community \cite{hu}. In
actual fact, recent experience suggests that unplanned emergent
reactor shutdowns would cause severe supply disruption. A number of
contingence incidents during last years has been pointing up
unreliability in the supply of medical radio-isotopes, particular
$^{99}\mbox{Mo}{/}^{99m}\mbox{Tc}$. Some $95\%$ of the world's
supply of these comes from only five reactors, all of them are over
$40$ years old \cite{hu}. So the greatest single threat to supply
reliability is the approaching obsolescence of the aging reactors
that current large-scale producers utilize to irradiate $HEU$
targets to elaborate the needful radio-isotopes. Last years, there
took place a number of significant disruptions in medical
radio-isotope supply, some of which have been lasting by now
\cite{int8,hu}. For instance, the concern about the long-term
supply of medical radio-isotopes has been exacerbated when the
shutdown of research reactor $HFR$ in the Netherlands since August
2008 has caused $^{99}\mbox{Mo}$ shortage world-wide. The most
productive and oldest, yet a while ago refurbished, Canada's $NRU$
reactor was shutdown in the summer 2009 \cite{nru}, after the heavy
water leak was discovered in May 15, 2009. It is not clear if and
when the $NRU$ could be restarted, or how to make up for its
outage. The worldwide supply of radio-isotopes is likely to be
unreliable unless newer production sources come on line.

Besides posing a threat to patients treatment, the current method
used by the world's main producers increases the menace of nuclear
terrorism, as it employs weapons-grade $HEU$. So the burning
question is now to eliminate, or at least minimize, the $HEU$ use
in reactor fuel, irradiated targets, and production facilities.
Only very few small-scale producers, {\it e.g.} in Argentina and
Australia, are, or are going to be able to manufacture
radio-isotopes using the $LEU$ targets \cite{hu}. The bulk of
consumed radio-isotopes are still obtained utilizing $HEU$ and, to
the best of knowledge, the conversion to the $LEU$ targets is not
believed before long. Especially the radio-isotope producing
community had been counting on the to-be-built reactors $MAPLE1$
and $MAPLE2$, and on the related processing facilities at
Chalk-River site, Canada, which would not have used the $HEU$. Yet
the $MAPLE$s, designed as a replacement for $NRU$, did not perform
as contemplated, and in May 2008 Atomic Energy of Canada Ltd. made
the decision to end the $MAPLE$'s project \cite{mapl}. This has, in
fact, put on hold any plans to convert to $LEU$-based large-scale
radio-isotope production. Instead, the world community these days
needs both new radio-isotope production inventions and the
facilities that will continue work safely in the long term, without
using weapon-grade uranium.

In this respect, we treat in what follows the photo-production of
various radio-isotopes, Sections 3, 4, which is due to the
bremsstrahlung induced in converter by initial electron beam of
electron accelerator, Section 2. Also we consider, in Section 5,
the case when a desirable isotope results in decaying a parent
radio-isotope that stems itself in the photo-production. At last,
in Section 6, the all-round discussion of findings persuades us
that the most preferable way to produce radio-isotopes is the usage
of electron beams provided by modern electron accelerators. What
encourages our work is the exploration by now carried out in the
Fefs. \cite{m1,m2,m3,mas3}.
\section*{2.
B\lowercase{remsstrahlung in converter}}
\label{sec:level2}
\setcounter{section}{2}
\setcounter{equation}{0}

As was proclaimed above, the purpose is to acquire how to work out
the various radio-isotopes, needful to-day for manifold
applications in technology, science and medicine, by making use of
electron beams delivered by microtrons, linear electron
accelerators, etc. That beam, with an electron energy distribution
$\rho_e(E_e)$ and a current density $J_e(t)[\mbox{A}/\mbox{cm}^2]$
(generally speaking, time dependent), travels through the converter
(see Fig. 1), which is prepared of some proper heavy element, such
as $W, \mbox{Pt}$ etc.The bremsstrahlung is thereby induced with
current density

\begin{equation}
J_{\gamma}(E_{\gamma})=\frac{{\cal N}_{\gamma}(E_{\gamma})}
{\mbox{s}{\cdot}\mbox{cm}^2{\cdot}\mbox{MeV}} \; ,\label{0i}
\end{equation}
expressed in terms of the photon number ${\cal
N}_{\gamma}(E_{\gamma})$ with the energy $E_{\gamma}=|{\bf k}|=k$,
per $1
\mbox{cm}^2,  1 \mbox{s}, 1 \, \mbox{MeV}$.

In turn, that $\gamma$-ray flux, interacting with respective nuclei
of the sample (see Fig. 1), induces the photo-nuclear reaction

\begin{equation}
\gamma + A(Z,N)\Longrightarrow A'(Z,N-1) + n , \label{00}
\end{equation}
so that a desirable isotope $A'(Z,N-1)$ comes out. Certainly, this
process (\ref{00}) can only be realized, if the energy $E_{\gamma}$
of $\gamma$ rays is, at least, greater than the neutron binding
energy $B_n$ of a considered nucleus $A(Z,N)$, ${E_{\gamma}}{>}{
B_n}{\approx}8 \, \mbox{MeV}$. Actually, the isotope $A(Z,N-1)$
production process will successfully run provided $E_{\gamma}$ is
of the order of, and comes over the energy $E_{GR}$ of giant
resonance in the photo-nuclear reactions (\ref{00}) on respective
nuclei, $E_{\gamma}\gtrsim E_{GR}(Z,N)\sim 13 \div 19
\mbox{\, MeV}$ \cite{nf}.
As a matter of course, an electron must have got the energy $E_e>
E_{\gamma}$ in order to give rise to the bremsstrahlung with the
required energy $E_{\gamma}$. Thus, only the processes involving
the electron and photon energies

\begin{equation}
\mbox{E}_{\gamma} \, , \, E_e \gtrsim E_{GR}\, \label{1}
\end{equation}
are to be taken into consideration and explored, which is the key
point of our treatment. Next, we limit the current study by the
condition

\begin{equation}
E_e\leq 100 \, \mbox{MeV} \label{ee}
\end{equation}
as well. The guide relations (\ref{1}), (\ref{ee}) govern all the
presented calculations, specifying the energy area where the
acquired findings hold true. Also, in the ordinary way, all the
evaluations we make in the work are the first $\alpha-$order, and
we abandon contributions from all the high $\alpha$-order
processes.

In passing across converter, a high energy electron is primarily
known to lose its energy (see e.c. Refs. \cite{ab,h,ll}) due to the
bremsstrahlung by scattering in the fields of nuclei of heavy atoms
of converter. As the relation (\ref{1}) holds, the angular
distribution of scattered electrons as well as emitted photons has
 got a sharp maximum in momentum direction of an initial electron.
Both electrons and photons spread within a small, rather negligible
solid angle $\Theta\sim { m }{/}{E_e}$ around direction of the
initial electron momentum \cite{ab,h,ll}. Then, with proper
allowance for screening, upon integrating the bremsstrahlung
cross-section over the angle between the momenta of incident
electron and emitted photon, a very handy expression for the
cross-section to describe the photon energy distribution results
(see {\it e.g.} Refs. \cite{h,sh,bh}):

\begin{eqnarray}
\frac{\mbox{d}\sigma_b(k)}{\mbox{d}k}=\frac{2 Z_C^2}{137}r_0^2\frac{1}
{k}\times \nonumber \\
\times {\{} \biggl(\frac{E_e^2+{E'_e}^2}{E_e^2}-
\frac{2E'_e}{3E_e}\biggr)\cdot\biggl(\ln M +1-\frac{2}{b}\arctan b
\biggr)+ \; \; \label{3} \\
+\frac{E'_e}{E_e}\biggl(\frac{2}{b^2}\ln(1+b^2)+
\frac{4(2-b^2)}{3b^3}\arctan b -\frac{8}{3b^2}+\frac{2}{9}\biggr)
{\}} \; , \nonumber
\end{eqnarray}
where $k=E_{\gamma}$ stands for the energy of radiated
$\gamma$-quantum, $E'_e=E_e-k$, $Z_C$ is the atomic number of the
converter material, and

$$b=\frac{2E_e E'_E Z^{1/3}}{C\, m
\, k} \, , \; \; \frac{1}{M}=\biggl(\frac{ m \, k} {2E_e
E'_e}\biggr)^2 +
\frac{Z_C^{2/3}}{C^2} \, , \; \; C=111 \, , \; \; r_0=\frac{e^2}{m}
=2.818\cdot 10^{-13}\, \mbox{cm}. $$
Besides the aforesaid bremsstrahlung in the field of nucleus,
 there exists the bremsstrahlung
  by scattering an incident electron by atomic electrons. For a
fast electron, $E_e\gg m$, the cross section of this process is
known to coincide with the bremsstrahlung cross section on nucleus
with $Z=1$ \cite{ab,h,ll}. Then, the atomic electrons contribution
into the whole electron bremsstrahlung is taken into account just
by replacing the factor $Z_C^2$ in Eq. (\ref{3}) by
$Z_C(Z_C+\delta)$ with $\delta\lesssim 1$. As for heavy converter
atoms $Z_C\gg 1$, this correction is rather of very small value.

The bremsstrahlung, with all the feasible energies $k=E_{\gamma}$,
causes the mean energy loss of electron on a unit of path
\cite{ab,h}

\begin{equation}
-\frac{\mbox{d}E_e(x)}{\mbox{d} x}={\cal N}_C\cdot E_e(x)\cdot
\varphi_{rad}(E_e) , \label{4}
\end{equation}
The number ${\cal N}_C$ of scattering atoms of converter in $1\,
\mbox{cm}^3$ is

\begin{equation}
{\cal N}_C = \frac{\rho_C\cdot6.022\cdot10^{23}}{A_C} , \label{5}
\end{equation}
where $\rho_C$ is the density of converter material, and $A_C$ is
its atomic weight. The quantity $\varphi_{rad}$ is written in the
form

\begin{equation}
\varphi_{rad}=\bar\varphi\cdot
K_C(E_e)=K_C(E_e) Z_C^2 \cdot 5.795\cdot10^{-28} \mbox{cm}^2
.\label{6}
\end{equation}
The coefficient $K_C$, very slightly varying with the energy $E_e$,
 provided $E_e\gtrsim 10\mbox{ \, MeV}$, can be found in Refs.
\cite{ab,h,gr} for various heavy atoms. So, upon passing a path
$x$, an electron with initial energy $E_e(0)$ will have got, in
consequence of the radiative losses, the energy

\begin{equation}
E_{e \; rad}(x)\approx E_e(0) \exp[-x{\cal N}_C\varphi_{rad}] .
\label{7}
\end{equation}

In fast electrons, $E_e\gg m $, elastic scattering on heavy nuclei
of converter, the angular distribution has got a very sharp
maximum, within the solid angle $\Theta<( m {/}E_e)^2$, and
therefore can be leaved out of our consideration \cite{ab,h,ll}.

In treating the fast electron collision with atomic electrons,
without photon emitting, we are to consider two cases. Firstly, let
the momentum $\Delta_I$ transferred to an atomic electron be

\begin{equation}
\Delta_I\lesssim I_Z\approx 13.5 Z_C \mbox{eV} , \label{8}
\end{equation}
the ionization potential of atom. Apparently, as $\Delta_I\ll E_e$,
a scattering angle is negligible. The mean electron energy loss on
a unit of path, caused by its inelastic collisions with atoms, is
described by the expression (see Refs. \cite{ab,h,bh})

\begin{equation}
-\frac{\mbox{d}E_{e}(x)}{\mbox{d}x}= 2\pi r_0^2 m {\cal N}_C
Z_C \ln\frac{E^3_{e}(x)}{2 m I_Z^2}
 \, ,\label{9}
\end{equation}
which can be rewritten in the form

\begin{equation}
x=-\frac{1}{6 \pi r_0^2 m {\cal N}_C Z_C}\int\limits_{E_e(0)}
^{E_{e \, I}(x)}\frac{\mbox{d}E}{\ln[E(2 m
 I_Z^2)^{-1/3}]} \, ,
\label{10}
\end{equation}
where $E_e(0)$ is the electron energy at the starting edge of
converter, and $E_{e \, I}(x)$ stands for the electron energy upon
passing the distance $x$, which is caused by the ionization losses.
With the conditions (\ref{1}), (\ref{ee}), we can actually presume

\begin{equation}
 \ln E \approx\ln E_e^{av} \, , \; \; \; \; E_e^{av}=\frac{E_e(0)
 +E_{GR}}{2} \, \label{dd}
\end{equation}
in the Eq. (\ref{10}). Then we arrive at the estimation of the
energy loss on the distance $x$ due to the inelastic electron
collisions with atoms

\begin{equation}
\Delta E_{e \, I}(x)\approx-x 6 \pi r_0^2 m {\cal N}_C Z_C
\ln[E_e^{av}(2  m  I_Z^2)^{-1/3}] \, . \label{11}
\end{equation}

Secondly, when, unlike (\ref{8}), the momentum transferred
$\Delta_I{\gg}I_Z$, yet still $\Delta_I{\ll}E_e$ anyway, atomic
electrons can be considered as free ones, and the fast electron
interaction with they reduces to the elastic forward scattering on
free resting electrons \cite{ab,ll}, which causes no energy loss,
as a matter of fact.

Amenably to Eqs. (\ref{7}), (\ref{10})-(\ref{11}), the electron
with the incident energy $E_e(0)$ at the starting edge of converter
 has got the energy

\begin{eqnarray}
E_e(x)\approx E_{e \, rad}(x)-\frac{ 6 \pi r_0^2 m Z_C
\ln[E_e^{av}(2  m  I_Z^2)^{-1/3}]}{\varphi_{rad}}\biggl(1-\frac
{E_{e \, rad}(x)}{E_e(0)}\biggr)\approx \nonumber\\
\approx E_{e \, rad}(x)+\Delta E_{e \, I}(x) \, , \label{12}
\end{eqnarray}
upon passing the path $x$ through converter (see Fig. 1). Just
this, $x$-dependent, energy $E_e(x)$ is to be substituted into Eq.
(\ref{3}) to describe the bremsstrahlung of an electron at the
distance $x$ from the starting edge of converter. Thus, the
bremsstrahlung production cross-section (\ref{3}) turns out to be
function of the distance $x$, via the electron energy $E_e(x)$
(\ref{12}).

In the actual evaluation explicated further in Sections 3, 4, the
converter thickness $R_C$ proves to be chosen so that there are no
electrons with the energies $E_e(R_C)\gtrsim 10 \mbox{\,MeV}
\gg m $ at the final edge of  converter.

 As expounded above, only the bremsstrahlung with $k\gtrsim 10
\mbox{\, MeV}
\gg  m $, described by Eq. (\ref{3}), is of value to induce the
desirable photo-nuclear reaction (\ref{00}). This bremsstrahlung,
caused by the initial electron beam with the energy distribution
$\rho_e(E_e)$ and the current density $J_e(t)$, when stems at a
distance $x$ from the starting edge of converter, is described by
the photon current density (\ref{0i})

\begin{eqnarray}
J_{\gamma}(x,k,E_e,Z_C,\rho_C,t)=\rho_e(E_e) J_e(t) {\cal N}_C
\frac{\mbox{d}\sigma_b(k,E_e(x),Z_C)}{\mbox{d}k} \, , \label{jg}
\end{eqnarray}
where the cross section ${\mbox{d}\sigma_b}{/}{\mbox{d}k}$ is given
by Eq. (\ref{3}) with the electron energy $E_e(x)$ (\ref{12}). This
 $\gamma$-flux spreads then forward, as was explicated above.

In this bremsstrahlung passing the path $(R_C - x)$ from a point
$x$ up to the final edge of converter $R_C$ (see Fig. 1), there are
three processes which cause the continuing $\gamma$-ray absorption
\cite{ab,h,ll} : 1) the $e^+ e^-$-pairs production; 2) the
photo-effect; 3) the Compton scattering on electrons, the first one
is known to be of the crucial importance at the considered
$k\gtrsim 10 \mbox{\, MeV}$ \cite{ab,h,ll}. Consequently, the
bremsstrahlung current density $J_{\gamma}(x,k)$ (\ref{jg})
decreases, becoming at the final edge of converter

\begin{eqnarray}
J_{\gamma}(x,k,R_C)=J_{\gamma}(x,k)\cdot\exp\biggl( - \frac{R_C-x}
{l_C(Z_C,{\cal N}_C,k,\rho_C)}\biggr) \, , \label{14}
\end{eqnarray}
where the length of absorption $l$ consists from three aforesaid
parts

\begin{equation}
\frac{1}{l_C}=
\frac{1}{l_{pair}}+\frac{1}{l_{photo}}+\frac{1}{l_{Com}} \, .
\label{15}
\end{equation}
Generally speaking, a tiny small quantity ${1}{/}{l_{\gamma \,
n}}$, caused by the reactions like (\ref{00}), should have been
added to right-hand side of Eq. (\ref{15}), for conscience's sake.
The values of $l$ for various materials are found, for instance, in
Refs. \cite{h,gr}. Let us mention that we deal with the
$\gamma$-ray energies just above the so-called ``area of maximum
transparency" \cite{ab,h,gr}.

As understood, precision of all the carried out calculations is
proved to be at least of the order $\sim\frac{m}{E_{GR}} \, , \;
\; \sim\frac{I_Z}{E_{GR}}$, that is anyway none the worse than
$\sim 10\%$.

With taken into consideration the restrictions imposed by the guide
conditions (\ref{1}), (\ref{ee}), we shall now discuss how the
cascade of electrons and photons, practicable to the isotope
production (\ref{00}), would emerge. The processes in those an
electron with the energy $E_e<E_{GR}$ participates can't anyway
lead to any discernible contribution into the photo-neutron
production
 (\ref{00}) of the desired isotope $A'(Z,N-1)$. In slowing-down
from the initial energy $E_e(0)$ to the energy $E_{GR}$, an
electron loses the energy

\begin{equation}
\tilde\Delta\approx E_e(0)-E_{GR} \, . \label{d}
\end{equation}
This energy loss $\tilde\Delta$ itself isn't considered to be
small. So, at the maxima currently treated electron energy
$E_e(0)=100 \, \mbox{MeV}$, we would have got
$\tilde\Delta\approx85 \, \mbox{MeV}$, and for the timely most
vital $E_e(0)=50 \, \mbox{MeV}$ we would arrive at
$\tilde\Delta\approx35 \, \mbox{MeV}$. As generally received
\cite{h,h1}, the primary share of this energy lost $\tilde\Delta$
is radiated most probably as the $\gamma-$rays with energies

\begin{equation}
\tilde k=\tilde E{\gamma}\approx\frac{\tilde\Delta}{2} \, .
\label{kk}
\end{equation}
Only a small part of this energy loss $\tilde\Delta$ is emitted as
a flux of comparatively soft photons, and $\gamma-$radiating with
the energies $k=E_{\gamma}>\tilde k$ proves to be all the more
negligible \cite{h,h1}. As was already discussed above, in
absorbing a photon with the considerable energy $\tilde k$
(\ref{kk}), the $e^+e^-$-pairs are produced with approximately
equal energies

\begin{equation}
E^+\approx E^-\approx\frac{\tilde k}{2}\approx\frac{\tilde\Delta}
{4} \, . \label{+-}
\end{equation}
Surely, there is no reason to suggest these energies to be as small
as negligible, yet anyway they are nevertheless substantially
smaller than the initial electron energy $E_e(0)$. Thus for the
timely most vital case $E_e(0)=50 \, \mbox{MeV}$, we have got
$E^{\pm}\approx 8 \, \mbox{MeV}<E_{GR}$, so that the thereby
produced $e^+ , \, e^- $ can never contribute to the isotope
production (\ref{00}) at all, which is understood in observing Fig.
2. Put another way, there would be a cascade, but the particles
participating therein would have got energies beyond the key
condition (\ref{1}). At the largest initial electron energy we
currently consider (\ref{ee}), $E_e(0)=100 \,
\mbox{MeV}$, there would be $E^{\pm}\approx 20 \, \mbox{MeV}$, so
as, generally speaking, these
 $e^+ , \, e^- $ themselves would give rise to the bremsstrahlung
 which could in turn serve to the isotope $A'(Z,N-1)$ photo-neutron
production (\ref{00}). Yet this isotope production, caused by those
secondary electrons with energies $E^{\pm}\approx 20 \,
\mbox{MeV}$, is anyway  10 times as small as the production due to
the initial electrons with $E_e(0)=100 \, \mbox{MeV}$ themselves,
which comes to light in observing the findings presented in tables
5,6,7, Section 4. Thus, when we abandon, even at $E_e(0)=100
\, \mbox{MeV}$, the above explicated cascade, the thereby inherent
ambiguouties will never come over $\approx 10\%$. That is why we do
not draw into consideration
 the bremsstrahlung which would be induced, in converter or in
sample, by the electrons those themselves would be originated by
absorption of the bremsstrahlung, which in its turn is due to
scattering an initial electron on nuclei in converter.

 Upon integrating Eq. (\ref{14}) over the initial electron energy
distribution and over the converter length, we obtain the
bremsstrahlung flux at the final edge of converter

\begin{eqnarray}
J_{\gamma \, C}(k,R_C,Z_C,\rho_C,E_E^b,E_e^u,\Delta_e,t)=
\int\limits_{E^b_e}^{E_e^u}\mbox{d}E\int\limits
_{0}^{R_C}\mbox{d}xJ_{\gamma}(x,k,R_C,E,Z_C,\rho_C,t) \, ,
\label{16}
\end{eqnarray}
where $E_e^b , \, E_e^u$ are, respectively, the bottom and upper
energies of the electron distribution in beam, and $\Delta_e$ is to
describe its width. In our further actual evaluations, the electron
energy varies between the limits $E_e^b , \, E_e^u$, and we choose

\begin{eqnarray}
\rho_e(E)=\frac{1}{n}\exp[-({(E-\bar E)}{/}{\Delta_e})^2] \, , \; \;
\bar E=\frac{E_e^b+E_e^u}{2} \, , \label{17}\\
1=\int \limits_{E_e^b}^{E_e^u}\mbox{d} E \rho_e(E) \, .\nonumber
\end{eqnarray}
Surely, when $\Delta_e\rightarrow 0$, Eq. (\ref{17}) reduces to the
$\delta$-function electron energy distribution. Also, let us recall
the initial electron current density $J_e$ in Eqs. (\ref{jg}),
(\ref{14}), (\ref{16}) is given in
${\mbox{A}}{/}{\mbox{cm}^2}={10^{19} \mbox{e}^-}{/} {(1.602\cdot
\mbox{sec}\cdot \mbox{cm}^2)}$
where $\mbox{e}^-$ is the electron electric charge.

\setcounter{section}{3}
\section*{3. R\lowercase{adio-isotope
photo-production in sample}}
\label{sec:level3}
\setcounter{equation}{0}

Traveling forward, the bremsstrahlung flux (\ref{16}) intrudes
 into the sample (see Fig. 1) that incorporates the isotope
$A(Z,N)$ which serves to produce the desirable radio-isotope
$A'(Z,N-1)$ due to the photo-nuclear reaction (\ref{00}). As well
known, this process is caused by the giant resonance in the nuclear
photo-absorption \cite{nf}. By now, there exist numerous reliable
measurements of the cross-sections $\sigma_{\gamma
\, n}(k,Z,N)$ of the neutron $\gamma$-production (\ref{00}) for
manifold nuclei. The respective data are put to use in our
evaluations. The errors in these $\sigma_{\gamma n}$ measurements
may rather amount $\sim 10\%$, which puts a bound to the accuracy
attainable in our evaluations. If anything, it is to point out that
at the large converted $\gamma-$ray energies, $E_{\gamma}\gtrsim 30
  \, \mbox{MeV}$, the contribution to (\ref{00}) from area beyond
 the giant-resonans nuclear photo-absorption would be discernible
because at these energies there exists
 the nuclear photo-absorption due to the surface absorption, the
virtual quasi-deuteron absorption, and the absorption caused by the
nucleon polarizability in nucleus \cite{246}. Though this
contribution ought to have been taken into account, its impact onto
the quantities we have been considering were hardly more than a few
per cent, even at $E_e(0)=100 \, \mbox{MeV}$.

 Absorption of the $\gamma$-flux goes on inside sample in much the
same way as in converter, yet $l_C$ (\ref{15}) gives place to
$l_S(Z_S,{\cal N}_S,\rho_S,k)$ of the sample. Upon passing a
distance $y$ from the starting edge of sample (see Fig. 1), the
$\gamma$-flux (\ref{16}) modifies as follows

\begin{eqnarray}
J_{\gamma \, S}(y,k,R_C,Z_C,\rho_C,E_e^b,E_e^u,\Delta_e,t)=
\nonumber \\
=J_{\gamma \, C}(k,R_C,Z_C,\rho_C,E_E^b,E_e^u,\Delta_e,t)
\exp[-\frac{y}{l_S(Z_S,{\cal N}_S,\rho_S,k)}]\, .\label{19}
\end{eqnarray}
Then the density of atoms ${\cal N}_{ris}(y,k,Z,N-1,t)$ of the
desirable radio-isotope $A'(Z,N-1)$, produced per $1 \mbox{s}$ by
the current density (\ref{19}), with a given $k$, at the distance
$y$ (see Fig.1) is determined by

\begin{eqnarray}
\frac{\mbox{d}{\cal N}_{ris}(y,k,Z,N-1,t)}{\mbox{d}t}=\nonumber\\
 ={\cal N}_S(Z,N)\cdot\sigma_{\gamma \, n} (k,Z,N)\cdot
J_{\gamma\,S}(y,k,R_C,Z_C,\rho_C,E_e^b,E_e^u,\Delta_e,t) \, ,
 \label{20}
\end{eqnarray}
where the density of sample atoms

\begin{eqnarray}
{\cal N}_S(Z,N)=\frac{\rho_S\cdot6.022{\cdot}10^{23}}{A_S} \,
\label{21}
\end{eqnarray}
is given in terms of the sample density $\rho_S$ and the atomic
 weight $A_S$. When the isotope $A(Z,N)$, needful to produce
 $A'(Z,N-1)$, constitutes only some part ${\cal A}_{bn}$ of the
 sample material, the density $\rho_S$ in (\ref{21}) is

\begin{equation}
\rho_S(Z,N)={\cal A}_{bn} \bar\rho_S \, , \label{p}
\end{equation}
where $\bar\rho_S$ is the whole sample density, in particular the
density of the
 natural element $A_S$. Unlike, the quantity $l_S$ in Eq.
(\ref{19}) is determined by the total density $\bar\rho_S$ anyway.

 Upon integrating the quantity (\ref{20}) over the length of sample
and over the photon energy $k$, we come to describe the total
amount of radio-isotope produced inside the sample, per 1 s, per
$1\mbox{cm}^2$ of a sample area,

\begin{eqnarray}
\frac{\mbox{d}{\cal N}_{ris}(Z,N-1,t)}{\mbox{d}t}
=J_e(t)\cdot{\cal N}_{ris}^0(Z,N-1) \, ,
\label{22} \\
{\cal N}_{ris}^0(Z,N-1)={\cal N}_S(Z,N)
\cdot{\cal N}_C(Z_C,N_C)\times \nonumber \\
\times\int\limits_{E_e^b}^{E_e^u}\mbox{d}E\rho_e(E)\int\limits_0
^{R_C}\mbox{d}x\int\limits_0^{\infty}\mbox{d}k\biggl(\frac{\mbox{d}
\sigma_b(k,E_e(x),Z_C)}{\mbox{d}k}\biggr)\sigma_{\gamma \,
n}(k,Z,N)\times \nonumber \\
\times\Biggl(1-\exp\biggl(-\frac{R_S}{l_S(k,Z_S,N_S)}\biggr)\Biggr)
\cdot l_S(k,Z_S,N_S)\times \nonumber \\
\times\exp\biggl(-\frac{R_C-x}{l_C(k,Z_C,N_C)}\biggr) \; .
\label{23}
\end{eqnarray}
The integration over the photon energy $k$ is actually restricted
by the area where the product

\begin{equation}
\biggl(\frac{\mbox{d}\sigma_b(k)}{\mbox{d}k}\biggr)\cdot\sigma_
{\gamma \, n}(k) \label{re}
\end{equation}
has got a discernible value. Beyond any questions, the values
$k\lesssim B_n$ and $k\gtrsim E_e$ contribute just nothing into
this integral over $k$ into the Eq. (\ref{23}).

The expression (\ref{22}) represents a source to produce this
isotope $A'(Z,N-1)$. To proceed further, we are to recall that the
produced radio-isotope $A'(Z,N-1)$ is not stable, and its decay is
governed by the life-time $\tau_s$, so that a number of decays per
$1\mbox{s}$ reads ordinarily
$$\frac{{\cal N}_{ris}(t,\tau_s)}{\tau_s}.$$
 Yet the isotope $A'(Z,N-1)$ itself undergoes irradiation by the
same $\gamma-$flux (\ref{19}) as the original isotope $A(Z,N)$
does. Then the photo-nuclear reaction

\begin{equation}
\gamma + A'(Z,N-1)\rightarrow A''(Z,N-2) + n \label{c1}
\end{equation}
results in depletion of the elaborated desired isotope $A'(Z,N-1)$,

\begin{equation}
-{\cal N}_{ris}(t,\tau_s)\frac{J_e(t) \, {\cal N}_{ris}^0(Z,N-1)}
{{\cal N}_S(Z,N)} \, .\label{c2}
\end{equation}
Then, amenably to the common equation

\begin{eqnarray}
\frac{\mbox{d}{\cal N}_{ris}(t,\tau_s)}{\mbox{d}t}=
J_e(t){\cal N}^0_{ris}(Z,N-1)-
\frac{{\cal N}_{ris}(t,\tau_s)}{\tau_s} - \nonumber\\
-{\cal N}_{ris}(t,\tau_s)
\frac{J_e(t) \, {\cal N}_{ris}^0(Z,N-1)}{{\cal N}_S(Z,N)}
 \, ,\label{24}
\end{eqnarray}
we obtain the radio-isotope amount, per $1 \mbox{cm}^2$ area of the
sample, elaborated during an exposition time $T_e$

\begin{eqnarray}
{\cal N}_{ris}(T_e,\tau_s)=
{\cal N}^0_{ris}
\int\limits_0^{T_e}\mbox{d}t J_e(t)\exp[{t}{/}{\tilde\tau_s}]
\cdot\exp[-{T_e}{/}{\tilde\tau_s}] \, ,
\label{25} \\
 \frac{1}{\tilde\tau_s
 \bigl(J_e(t)\bigr)}{=}\frac{1}{\tau_s}{+}\frac{J_e(t)
\, {\cal N}_{ris}^0(Z,N-1)}{{\cal N}_S(Z,N)} \, .\label{c5}
\end{eqnarray}
Although, strictly speaking, the cross-section $\sigma_{\gamma
n}(k,Z,N)$ in the expression ${\cal N}^0_{ris}$ in Eq. (\ref{c2}),
and in the last terms in Eqs. (\ref{24}), (\ref{c5}) would give
place to $\sigma_{\gamma n}(k,Z,N-1)$, we utilize here
$\sigma_{\gamma n}(k,Z,N-1)\approx\sigma_{\gamma n}(k,Z,N)$ in the
 evaluations of these correction terms. For a time-independent
initial electron current $J_e$, Eq. (\ref{25}) reduces to

\begin{eqnarray}
{\cal N}_{ris}(T_e,\tau_s)= {\cal N}^0_{ris} J_e \tilde
\tau_s (1-\exp[-{T_e}{/}{\tilde\tau_s}]) \, . \label{26}
\end{eqnarray}
When $T_e\ll\tau_s$, this gets simplify, giving just

\begin{equation}
{\cal N}_{ris}(T_e,\tau_s)= {\cal N}^0_{ris}J_e T_e \, . \label{27}
\end{equation}
Let us mention that though the correction (\ref{c2}) is to be
allowed for, its impact on the isotope production is very small,
rather negligible, at the values of $T_e , \, J_e$ currently
treated.

 It is to designate that we have been using, all over the carried
out calculations, just the life-time $\tau$, yet not the so-called
half-decay period $T_{{1}{/}{2}}=\tau\ln 2$.

It is generally accepted (see, for instance, Refs.
\cite{ir,m1,m2,m3,mas3}) to describe the radio-isotope production
in terms of the yield $Y[{\mbox{Bq}}{/}{\bigl(\mbox{h}\cdot
\mu\mbox{A}\cdot\mbox{mg}{A(Z,N)}\bigr)}]$
of the produced activity in $\mbox{Bq}$ per $1 \mbox{h}$ of
exposition time, per $1\mu\mbox{A}$ of the initial electron
current, and per $1\mbox{mg}$ of the isotope $A(Z,N)$ in the
sample, which serves to produce the desirable isotope $A'(Z,N-1)$.
Accordingly its definition, this characteristic $Y$ is expressed
through the quantity (\ref{25})

\begin{eqnarray}
Y=\frac{{\cal N}_{ris}(Z,N-1,J_e,E_e^b,E_e^u,R_C,R_S,{\cal
A}_{bn},T_e,\tau_s)}
{R_S(\mbox{cm})\cdot\rho_S({\mbox{mg}}{/}{\mbox{cm}^3})A(Z,N)\cdot
\tau_s(\mbox{s})\cdot T_e(\mbox{h})\cdot J_e(\mu\mbox{A})}
 \, . \label{29}
\end{eqnarray}
It is also of use to discuss the total yield of activity produced
by the initial electron current $J_e$ inside the whole actual
sample, with $1\mbox{cm}^2$ area and thickness $R_S$, during
exposition time $T_e$

\begin{eqnarray}
{\cal Y}(\mbox{Bq})=Y\cdot
R_S(\mbox{cm})\cdot\bar\rho_S({\mbox{mg}} {/}{\mbox{cm}^3})\cdot
{\cal A}_{bn}\cdot
 J_e(\mu\mbox{A})\cdot T_e(\mbox{h}) \, . \label{30}
\end{eqnarray}
Beside $Y\, , \; \, {\cal Y}$ (\ref{29}), (\ref{30}), it is of
value to consider the total amount of radio-isotope $A'(Z,N-1)$
elaborated in the whole sample

\begin{equation}
{\cal M}_{Z,N-1}(\mbox{g})=\frac{{\cal
N}_{ris}(Z,N-1,J_e,E_e^b,E_e^u,R_C,R_S,{\cal A}_{bn},T_e,\tau_s)
\cdot {\cal A}_{bn}\cdot(Z+N-1)}{6.022{\cdot}10^{23}} \, , \label{31}
\end{equation}
where $(Z+N-1)=A-1$ is the corresponding atomic weight. In Sections
4 and 5, we display the results of $Y$ and ${\cal M}$ evaluations.
Apparently, a ${\cal Y}$ value is directly expressed through a $Y$
value accordingly (\ref{30}).

Surely, besides the desirable isotope $A'(Z,N-1)$ photo-production
(\ref{00}), the reactions

\begin{eqnarray}
A(Z,N)+\gamma=A'(Z,N-2)+2n \, , \label{i}\\
A(Z,N)+\gamma=A'(Z-1,N)+p \, , \label{ii} \\
A(Z,N)+\gamma=A'(Z-1,N-1)+p+n \, \label{iii}
\end{eqnarray}
are generally known to take place as well. Yet their thresholds are
 nearly twice as much as the threshold of the reaction (\ref{00}),
and their cross-sections are about ten times as small as the
cross-section of the reaction (\ref{00}) \cite{nf}. So, with
accuracy quite sufficient, the processes (\ref{i}), (\ref{ii}),
(\ref{iii}) are not competitive with the considered main
photo-production (\ref{00}) of the isotope $A'(Z,N-1)$. Surely,
when desired, the
 yield of isotopes $A'(Z,N-2),\, A'(Z-1,N-1), \, A'(Z-1,n)$ from
 irradiated sample would be calculated as well.

Let us recall that the eventual results (\ref{25})-(\ref{31}) are
governed by the manifold parameters, which characterize 1) the
initial electron beam, $J_e, T_e, \bar E_e, E_e^b, E_e^u,
\Delta_e$; 2) the converter, $\sigma_b, {\cal N}_C, \rho_C, R_C,
l_C, K_C$; 3) the sample and the produced radio-isotope,
$\sigma_{\gamma
\, n}, {\cal N}_S, \rho_S,$
 $Z_S, N_S, {\cal A}_{bn}, l_S, R_S, N, Z, \tau_s$. In involving
these quantities into consideration,
 the proper discussions were explicated above. The dependence of
$Y, {\cal M}, {\cal Y}$ on these parameters will be considered in
next Section 4 for some radio-isotopes, produced immediately in the
reaction (\ref{00}). Thereafter, in Section 5, we inquire into the
event that the decay

\begin{equation}
A'(Z,N-1) \Longrightarrow A^m (Z',N')
\end{equation} of this, at the first step obtained radio-isotope
$A'(Z,N-1)$, serves, in turn, as a source to produce the
second-step radio-isotope $A^m$, which is eventually put to use in
manufacturing the needful practicable preparation.

\section*{4. E\lowercase{valuation of the radio-isotope
production characteristics}}
\label{sec:level4}
\setcounter{section}{4}
\setcounter{equation}{0}
Now we evaluate the quantities $Y, {\cal M}, {\cal Y}$, acquired
above, for some examples, which typify the radio-istope production
using the electron beam. All over further consideration, the
converter is presumed to be prepared of $^{nat}W$ (the natural
tungsten) with a varying thickness $R_W (\mbox{cm})$, and three
cases will be considered: the production of $^{99}\mbox{Mo} \, ,\;
\; ^{237}\mbox{U} \, , \; \; ^{117m}\mbox{Sn}$, with varying the
sample thickness $R_{\mbox{Mo}} \, , \;
\; R_{\mbox{U}} \, , \; \; R_{\mbox{Sn}}$, the
initial electron energy $E_e$ and the current $J_e$, and the
exposition time $T_e$. It is implied that we have been dealing with
the average electron current provided by the electron linear
accelerator or microtron. The purpose is to visualize the
dependence of the production characteristics $Y, {\cal M}, {\cal
Y}$ (\ref{29})-(\ref{31}) on the aforesaid parameters.

First of all we treat the reaction

\begin{equation}
\gamma + \, ^{100}\mbox{Mo}\Longrightarrow \, ^{99}\mbox{Mo}
 + n \label{32}
\end{equation}
providing the production of the isotope $^{99}\mbox{Mo}$, which is
known to be the most applicable \cite{int7,hu}, as discussed in
Introduction. Let us recall the $^{99}\mbox{Mo}$ life-time
$\tau_{^{99}\mbox{Mo}}\approx 96 \mbox{h}.$

As indicated by Eqs. (\ref{i}), (\ref{ii}), (\ref{iii}), the
isotopes $^{98}\mbox{Mo} , \, ^{98}\mbox{Nb} , \, ^{99}\mbox{Nb}$
could be recovered as well, if required.

In order to elucidate the key point of treatment, we display in
Fig. 2. the $\gamma$-flux $J_{\gamma \, C}(k)$ (\ref{16}) at the
various thickness $R_{\mbox{W}}$ of $\mbox{W}$-converter and at the
various initial electron energy (\ref{17}). The behavior of
$J_{\gamma
\, C}(k)$ is to be correlated with the energy dependence of the
cross-section of the reaction (\ref{32}) \cite{sm}. As understood,
only the area of $k$-values where $J_{\gamma \, C}$ and
$\sigma_{\gamma \, n}$ overlap determines the evaluated quantities
(\ref{22})-(\ref{31}) to describe the radio-isotope
$^{99}\mbox{Mo}$ production, which we are treating now. The
$\gamma$-rays with energies $k$ beyond this area are out of value.
As the function
 $\sigma_{\gamma \, n}$ is well known to be, more or less, of the
same form and magnitude for all the heavy and middle weight nuclei,
 Fig. 2 typifies the calculation of radio-isotope production by
means of electron beam.

As explicated in Section 2, the simultaneous treatment of botch
  bremsstrahlung production and absorption and electron energy
losses serves to realize how the isotope yield does depend on the
converter thickness $R_{\mbox{W}}$. This dependence is typified by
 table 1. As understood there exists the most preferable
$R_{\mbox{W}}$-value for given material of the converter and the
incident electron energy. In actual radio-isotope manufacturing,
just this $R_C$ is to be utilized. Let us mention that the quantity
$Y{\approx}3.2{\,
\mbox{kBq}}{/}{(\mbox{h}\,{\mu}\mbox{A mg}^{100}\mbox{Mo})}$ was
obtained in Ref. \cite{m1} at
 $R_{\mbox{W}}=0.3 \mbox{cm}$, which is some greater than the most
preferable value $R_{\mbox{W}}\approx0.17\mbox{cm}$. As seen, the
result of $Y$ measurement in Ref. \cite{m1} does actually coincide
with ours in table 1, with a reasonable accuracy.

The dependence of $Y, {\cal M}$ on the $R_S$ value at the various
electron energies $E_e$ gets understood from the data given in the
tables 2, 3, 4. This $Y \,, \; {\cal M}$ behaviour is due to
simultaneous run of the isotope $^{99}\mbox{Mo}$ photo-production
and the $\gamma$-rays absorption in the $\mbox{Mo}$ sample, as was
acquired in Section 3. With $R_{\mbox{Mo}}$ growing, the quantity
${\cal N}_{ris}$ (\ref{22}), (\ref{25}) increases substantially
slower than linearly. That is why the quantity $Y$ decreases, and
${\cal M}$ tends to some limit by increasing $R_{\mbox{Mo}}$.

The quantity ${\cal Y}$ (\ref{30}) behaves rather alike ${\cal M}$.
With the same parameters, as in table 3, for the natural
$^{nat}\mbox{Mo}$ sample, $\rho_{\mbox{Mo}}={9,8\mbox{g}}
{/}{\mbox{cm}^3}, {\cal A}_{bn}=0.1$, with the thickness
$R_{\mbox{Mo}}=2\mbox{cm}$ and $1\mbox{cm}^2$ area, during
$1\mbox{h}$ exposition time, we have got

\begin{equation}
{\cal Y}_{^{99}\mbox{Mo}} \approx 1.7\cdot 10^{10}{\mbox{kBq}}
\, ,
\label{33}
\end{equation}
which is rather noteworthy. If anything, by dividing the yield
(\ref{30}) on total sample mass, one might treat the so-called
specific activity ${\cal Y}_{^{99}\mbox{Mo}}^{sp}$, which measures
the activity of $^{99}\mbox{Mo}$ per unit mass of reaction
products,

\begin{equation}
{\cal Y}_{^{99}\mbox{Mo}}^{sp}=
\frac{{\cal Y}_{^{99}\mbox{Mo}}}{\rho_{^{99}\mbox{Mo}}
\cdot R_{^{99}\mbox{Mo}}\cdot 1 \mbox{cm}^2}
\approx 10^9\frac{\mbox{kBq}}{\mbox{g}}.
\label{yyy}
\end{equation}
Yet this quantity would be rather of small use, as it does actually
depend on the sample and converter parameters, $R_C, R_S, Z_C, Z_S,
\rho_C, \rho_S$, the exposition time $T_e$,
the percetage ${\cal A}_{bn}$ of $^{100}\mbox{Mo}$ in the sample,
and so on. That is why we have been making use of ${\cal
Y}_{^{99}\mbox{Mo}}$ itself without having recourse to ${\cal
Y}_{^{99}\mbox{Mo}}^{sp}$ (\ref{yyy}).

Fig. 3 offers the time dependence of the quantities $Y$ (\ref{29})
and ${\cal M}$ (\ref{31}). They represent the produced activity and
mass as a function of exposition time $T_e$ in hours. With $T_e$
increasing, the quantity ${\cal N}_{ris}$ grows tangibly slower
than linearly. Consequently, $Y(T_e)$ decreases and ${\cal M}(T_e)$
tends to a finite limit when $T_e$ increases. As seen, there is no
reason for too long exposition time.

Table 5 demonstrates how the quantities $R_{\mbox{W}} \, , \; Y$
and ${\cal M}$ depend on the initial electron energy. As seen, the
most preferable $R_{\mbox{W}}$
 value increases smoothly with electron energy growth. The feature
to emphasize is the sharp augmentation of $Y\, , \; {\cal M}$ in
coming from $\bar E_e=20 \mbox{\, MeV}$ to $\bar E_e=25
\mbox{\, MeV}$ and then to $\bar E_e=50 \mbox{\, MeV}$,
yet the relatively slower enhancement shows up by increasing $\bar
E_e$ from $\bar E_e=50 \mbox{\, MeV}$ to $\bar E_e=100 \mbox{\,
MeV}$.

The next example of considerable practical interest \cite{s} is the
photo-production of the tin isomer radio-isotope $^{117m}\mbox{Sn}$

\begin{equation}
\gamma + \, ^{118}\mbox{Sn} \Longrightarrow \,
^{117m}\mbox{Sn} + n \, . \label{34}
\end{equation}

We put to use the cross-section of this reaction acquired from Ref.
\cite{ss}. The results of $Y$ (\ref{29}) and ${\cal M}$ (\ref{31})
evaluation are presented in table 6 for various initial electron
energies and for various thicknesses $R_{\mbox{W}}$ and
$R_{\mbox{Sn}}$ of the $\mbox{W}$ converter and of the natural tin
$^{nat}\mbox{Sn}$ sample, with $1\mbox{cm}^2$ area. Let us recall
that the isotope $^{118}\mbox{Sn}$ constitutes about $24\%$ of the
natural tin, that is ${\cal A}_{bn}\approx0.24$ in (\ref{p}),
$\bar\rho_{\mbox{Sn}}=7 {\mbox{g}}{/}{\mbox{cm}^3}$, and the isomer
$^{117m}\mbox{Sn}$ life time $\tau_{^{117m}\mbox{Sn}}\approx 20.2
\mbox{d}$ \cite{gr}. As seen, this table 6 offers the same
dependence of $Y$ (\ref{29}) and ${\cal M}$ (\ref{31}) on the
electron energy for the $^{117m}\mbox{Sn}$ production, as the table
5 for the production of $^{99}\mbox{Mo}$ does. Accordingly to the
data in this table 6, the quantity ${\cal Y}$ (\ref{30}) at $\bar
E_e=50
\mbox{\, MeV}
\, , \; \; R_{\mbox{Sn}}=2\mbox{cm}$ proves to be

\begin{equation}
{\cal Y}_{^{117m}\mbox{Sn}}\approx 0.8\cdot 10^{10}\mbox{kBq} \, ,
\label{35}
\end{equation}
which is of the same order, as the ${\cal Y}$ (\ref{33}) for
$^{99}\mbox{Mo}$ production.

At last, we discuss the production of the widely applied \cite{u},
specifically in the nuclear fuel research, radio-isotope
$^{237}\mbox{U}$,

\begin{equation}
\gamma + \, ^{238}\mbox{U} \Longrightarrow \,
^{237}\mbox{U} + n \, .\label{36}
\end{equation}
The cross section of this reaction is acquired from Ref. \cite{us}.
All the consideration runs in much the same way as in the cases of
treating the $^{99}\mbox{Mo}$ and $^{117m}\mbox{Sn}$ production.
Yet now we evaluate the quantities $Y$ (\ref{29}), ${\cal M}$
(\ref{31}) not just at the time $T_e$, the finish of the
exposition, but in one day after the $5\mbox{h}$-long irradiation.
The reason to do so is that the experimental measurements of the
$^{237}\mbox{U}$ production were carried out in Ref. \cite{m2} just
under such conditions. Apparently, as the time of observation $T$,
counting from the start of irradiation, is longer than the
exposition time $T_e$, the quantities (\ref{29}), (\ref{31}) are to
be replaced by

\begin{eqnarray}
Y(T)=Y(T_e) \exp[-{(T-T_e)}{/}{\tau_{^{237}\mbox{U}}}] \,
,\label{37} \\ {\cal M}(T)={\cal M}(T_e)
\exp[-{(T-T_e)}{/}{\tau_{^{237}\mbox{U}}}] \, ,\label{38}
\end{eqnarray}
with the $^{237}\mbox{U}$ life time $\tau_{^{237}\mbox{U}}=6.75
\mbox{d} \, , \; T-T_e=1 \mbox{d} $,
which are presented in table 7. For the parameters utilized in Ref.
\cite{m2}, $R_{\mbox{W}}=0.3 \mbox{cm}\, , \; \, \bar E_e=24
\, \, \mbox{MeV}\, , \; \, \Delta_e\rightarrow 0 $, we have got
$Y_{\mbox U}{\approx}1.35\, {\mbox{kBq}}{/}{(\mbox{h}\,
\mu\mbox{A}\,
 \mbox{mg} ^{238}\mbox{U})}$, which is in accordance with the
result obtained in \cite{m2},
 $Y_{\mbox U}{\approx}1.1\, {\mbox{kBq}}{/}{(\mbox{h}\,
\mu\mbox{A}\, \mbox{mg} ^{238}\mbox{U})}$. Let us recall the
isotope $^{238}\mbox{U}$ constitutes $99.276\%$ of the natural
$\mbox{U}$, that is ${\cal A}_{bn}{\approx}1$ in (\ref{p}),
(\ref{30}), (\ref{31}), and
$\rho_{\mbox{U}}{=}18.7{\mbox{g}}{/}{\mbox{cm}^3}$. Then, utilizing
the data from table 7, and choosing $\bar E_e =50 \mbox{\, MeV}
\, , \; R_{\mbox{U}}
=2\mbox{cm} \, , \; T_e =5 \mbox{h}$
(unlike the cases of $^{99}\mbox{Mo} \, , \;
^{177m}\mbox{Sn}$), $T=30 \mbox{h}$, the quantity ${\cal Y
\,}$ (\ref{30}) proves to be

\begin{equation}
{\cal Y}_{^{237}\mbox{U}}\approx35\cdot 10^{10} \mbox{kBq} \, ,
\label{u} \end{equation}
that is still more significant than the ${\cal Y}$ values for
$^{99}\mbox{Mo}$ and $^{117}\mbox{Sn}$, (\ref{33}), (\ref{35}).

\section*{5. P\lowercase{roduction of a practicable
radio-isotope via an isotope-precursor}}
\label{sec:level5}
\setcounter{section}{5}
\setcounter{equation}{0}

In a number of important cases, a radio-isotope of practical use
$A^m(Z',N')$ is generated by decaying

\begin{equation}
A'(Z,N-1)\Longrightarrow A^m(Z',N') + e^{\pm} \label{gen}
\end{equation}
an isotope $A'(Z,N-1)$ obtained in the photo-nuclear reaction
(\ref{00}), which was described in previous Section 4. Thus, the
parent isotope decay (\ref{gen}) is now a source to produce
 a needful eventual radio-isotope $A^m(Z',N')$ with a life-time
$\tau_m$. The density of atoms ${\cal N}_{ris}^m$ of this isotope
$A^m$ produced by the $\gamma$-flux (\ref{19}) inside a given
sample is then described by the common equation

\begin{equation}
\frac{\mbox{d}{\cal N}_{ris}^m(t,\tau_s,\tau_m)}{\mbox{d}t}=
p_m\frac{{\cal N}_{ris}(t,\tau_s)}{\tau_s} -
 \frac{{\cal N}_{ris}^m(t,\tau_s,\tau_m)}{\tau_m} \,. \label{40}
\end{equation}
Here $p_m$ stands to allow for the fact that the isotope decay
(\ref{gen}) constitutes a share $p_m$ of all the possible decays of
$A(Z,N-1)$. Amenably Eq. (\ref{40}) we generally obtain the amount
of radio-isotope $A^m (Z',N')$ for the sample with $1\mbox{cm}^2$
area, at the total time T, counting from the exposition start,

\begin{eqnarray}
{\cal N}_{ris}^m(T,\tau_s,\tau_m)=p_m
\frac{\exp[-{T}{/}{\tau_m}]}{\tau_s}
\int\limits_0^T\mbox{d}t\exp[{t}{/}{\tau_{m}}]
{\cal N}_{ris}(t,\tau_s) \, , \label{41}
\end{eqnarray}
with the quantity ${\cal N}_{ris}(t,\tau_s)$ given by Eqs.
(\ref{24})-(\ref{27}). Ordinarily, we are dealing with the
practicable case $T\geq T_e$. In calculating ${\cal N}^m_{ris}$
(\ref{41}), we have abandoned leaking the isotope $A^m(Z',N')$ due
to the feasible photo-nuclear reaction
$A^m(Z',N')(\gamma,n)A''(Z',N'-1)$ during exposition time $T_e$. In
the case a constant initial electron current $J_e$ irradiates
converter during the exposition time $T_e$, and afterwards
disappears, the general Eq. (\ref{41}) reduces to

\begin{eqnarray}
{\cal N}_{ris}^m(T,T_e,\tau_s,\tau_m)={\cal N}_{ris}^0 J_e p_m
\frac{\tilde\tau_s}{\tau_s}
\Biggl(\exp[-{T}{/}{\tau_m}]\biggl(\tau_m\Bigl(\exp[{T_e}{/}
{\tau_m}]-1\Bigr)-\nonumber\\
-\tau_{-}\Bigl(\exp[{T_e}{/}{\tau_{-}}]-1\Bigr)+
\tau_{-}\Bigl(1-\exp[-{T_e}{/}{\tilde\tau_s}]\Bigr)\Bigl(\exp[{T}{/}
{\tau_{-}}]-\exp[{T_e}{/}{\tau_{-}}]\Bigr)\biggr)\Biggr) \, ,
\label{42}\\
\frac{1}{\tau_{-}}=\frac{1}{\tau_m}-\frac{1}{\tilde\tau_s} \, .
\; \; \; \;  \; \; \; \nonumber
\end{eqnarray}
With replacing the quantity ${\cal N}_{ris}(T_e,\tau_s)$ in Eqs.
(\ref{29})-(\ref{31}) by ${\cal N}_{ris}^m(T,T_e,\tau_s,\tau_m)$
(\ref{41}), (\ref{42}), and $\tau_s$ by $\tau_m$ as well, the
quantities $Y^m$ and ${\cal M}^m$ are defined to describe the
eventual isotope $A^m(Z',N')$ production.

Hereafter, using the Eq. (\ref{42}), we treat the production of the
most extensively employed radio-isotope $^{99m}\mbox{Tc}$
\cite{int7,hu}, which stems in the $^{99}\mbox{Mo}$ $\beta-$decay

\begin{equation}
^{99}\mbox{Mo} \longrightarrow \, ^{99m}\mbox{Tc} +
 e^- \, . \label{43}
\end{equation}
The branching of the $\beta-$decay of $^{99}\mbox{Mo}$ into the
isomer $^{99m}\mbox{Tc}$, with the life-time $\tau_m{\approx}10
 \mbox{h}$ , amounts $\approx 85\%$, that is $p_m\approx 0.85$ in
Eq. (\ref{42}) \cite{ttc}. In other cases $^{99}\mbox{Mo}$ decays
giving the practically stable $^{99}\mbox{\small Tc}$ isotope with
$\tau_{\mbox{Tc}}\approx 4\cdot
 10^5 \mbox{y}$ \cite{gr,m1,ttc}. All the results presented
hereafter are obtained for the $\mbox{Mo}$ sample of $1
\mbox{cm}^2$ area, as had been doing in  previous Sections 2, 3,
4.

 Tables 8, 9 show that for a given $T_e$ there exists the most
preferable time $T_{max}$, counting from the exposition start, when
the yield of $^{99m}\mbox{Tc}$ activity $Y_{\mbox{Tc}^m}$ and the
mass ${\cal M}_{\mbox{Tc}^m}$ have got their maxima. Therefore the
radio-isotope $^{99m}\mbox{Tc}$ is to be extracted out of the
$\mbox{Mo}$ sample upon the time $T_{max}$ best of all. This fact
is thought to be of a practical value, though the dependence of
$Y_{\mbox{Tc}^m}$ and ${\cal M}_{\mbox{Tc}^m}$ yield on $T$ sees to
be rather smooth.

 Table 10 shows the activity yield $Y_{\mbox{Tc}^m}$ decreases and
the mass ${\cal M}_{\mbox{Tc}^m}$ tends to a certain limit as the
thickness $R_{\mbox{Mo}}$ of $\mbox{Mo}$ sample increases. That is
so because the quantity ${\cal N}_{ris}^m$ (\ref{41}) increases
substantially slower than linearly, just alike the quantity ${\cal
N}_{ris}$ (\ref{25}) does.

 Tables 11, 12 figure the quantities $Y_{\mbox{Tc}^m}(T)$, ${\cal
M}_{\mbox{Tc}^m}(T)$ calculated at $T=T_e$ and $T=T_{max}$, for the
various initial electron energies $\bar E_e$ (attached by the
associated parameters $E_e^b \, , \; E_e^u \, , \; \Delta_e$ as in
table 5), with choosing the most preferable values of the converter
thickness $R_{\mbox{W}}$ and the time $T_{max}$ for each $\bar
E_e$. It is of interest to correlate the results obtained for the
$\mbox{Mo}$ foil, $R_{\mbox{Mo}}=0.01
\mbox{cm}$, with those for the thick enough $\mbox{Mo}$ sample,
$R_{\mbox{Mo}}=2\mbox{cm}$. Just for that matter,
 the evaluation firstly carried out at $R_{\mbox{Mo}}=0.01
 \mbox{cm}$ is then replicated at $R_{\mbox{Mo}}=2 \mbox{cm}$, with
 the results offered in tables 11 and 12, respectively.

Accordingly the data in table 12, the radio-isotope production
characteristic ${\cal Y}$ (\ref{30}), the total yield of
$^{99m}\mbox{Tc}$ activity,
 at $\bar E_e=50
\mbox{\,MeV}$ is found to be

\begin{equation}
{\cal Y}_{\mbox{Tc}^m}(T_{max})\approx 1.2\cdot 10^{10}
\mbox{kBq}
\, ,\label{44}
\end{equation}
that is of the same order as the corresponding value (\ref{33}) for
$^{99}\mbox{Mo}$. In other words, let one irradiate the natural
$^{nat}\mbox{Mo}$ sample, with the $1\mbox{cm}^2$ area and the
thickness $R_{\mbox{Mo}}=2\mbox{cm}$, during $T_e=1\mbox{h}$ by the
$\gamma$-flux originated by the electron beam, with $\bar E_e=50
  \, \mbox{MeV}$ and $J_e={1 \mbox{A}}{/}{\mbox{cm}^2}$, in the
$^{nat}W$ converter with the thickness $R_{\mbox{W}}=0.3
\mbox{cm}$. Then, the $^{99}\mbox{Mo}$ activity (\ref{33}) would
be elaborated by the end of exposition, and, in turn, the
$^{99m}\mbox{Tc}$ activity (\ref{44}), comparable with (\ref{33}),
would be generated at the time $T_{max}=23.5 \mbox{h}$

It is to mention that upon extracting the isotope $^{99m}\mbox{Tc}$
out of the $\mbox{Mo}$ sample at $T=T_{max}$, the next amount of
$^{99m}\mbox{Tc}$ isotope, comparable with that at $T=T_{max}$,
would be accumulated in the $\mbox{Mo}$ sample in
$T_1\approx\tau_{^{99m}\mbox{Tc}}\ln({\tau_{^{99}\mbox{Mo}}}{/}
{\tau_{^{99m}\mbox{Tc}}})
\approx23\mbox{h}$, as can be realized from Eq.
(\ref{42}). Yet further repeatedly, over and over again,
withdrawing $^{99m}\mbox{Tc}$ out of the sample is usually thought
to be rather of less efficiency because of diminishing the
$^{99}\mbox{Mo}$ amount.

Of course, the examples treated in the presented work do not cover
all the area of the radio-isotope application in these days. In
particular, there exists a number of practicable isotopes eligible
to be wrought up by the method described hereupon, in the
photo-nuclear reactions such as

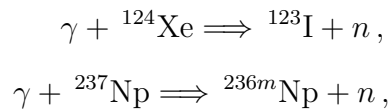
\begin{eqnarray}
\gamma+\,^{124}\mbox{Xe} \Longrightarrow\,^{123}\mbox{I}+n \,
,\nonumber\\
\gamma+\,^{237}\mbox{Np} \Longrightarrow\,^{236m}\mbox{Np}+n \,
,\nonumber
\end{eqnarray}
and so on \cite{m3,isi}. Further investigations in this way are
believed to be carried out before long.

Aforesaid obtaining $^{99m}\mbox{Tc}$ out of $^{99}\mbox{Mo}$
typifies manufacturing the practicable radio-isotope due to decay
of some preceding isotope, procured at the first step, in the
photo-nuclear reaction (\ref{00}).

The values of $Y \,, \; {\cal Y} \, , \; {\cal M}$ acquired in
Secs. 4, 5 are believed to be of practical interest, which is
discussed next. Let us here recollect that accuracy of our findings
is at all the points about $\sim 10\%$, as was explicated at every
stage in carrying out the presented calculations.

\section*{6. F\lowercase{indings consideration}}
\label{sec:level}
\setcounter{section}{6}
\setcounter{equation}{0}
Once, for all we have by now acquired, there emerge alluring
prospects of the radio-isotopes photo-production around electron
linear accelerators.

Hereafter we correlate and contrast salient features of the routine
reactor-based radio-isotope production, even though with the
conversion from the $HEU$- to $LEU$-targets, and the electron
accelerator-based radio-isotope production, addressing advantages
of the last, specifically with respect to the case of recovery of
$^{99}\mbox{Mo}$ and $^{99m}\mbox{Tc}$, which are by far the major
medical isotopes salable and consumed to-day
\cite{int1,int2,int7,hu}. We follow the timely industry convention,
and quantify the radio-isotope production and supply in terms of
{\it 6-day curies per week} \cite{int1,int2,hu}, which is nominally
the quantity of $^{99}\mbox{Mo}$ activity remaining 6 days after
the recovered $^{99}\mbox{Mo}$ leaves the producer's facility,
provided the $^{99}\mbox{Mo}$ has been elaborated during one week,
and then was refined and processed before shipment to the market.

Let the electron beam with $J[\mbox{A}{/}{\mbox{cm}^2}]$ and $\bar
E_e=50 \, \mbox{MeV}$ ( see Fig. 1) irradiate the tungsten
converter with $R_{\mbox{W}}=0.3\mbox{cm}$, and the $\gamma-$flux,
converted from this electron beam, produce the isotope
 $\mbox{Mo}^{99}$ in the $\mbox{Mo}$ sample, with
$R_{\mbox{Mo}}=2\mbox{cm}$ and $1 \mbox{cm}^2$ area, amenably to
the photo-nuclear reaction (\ref{32}). Then, with allowance for the
time-dependence of $Y$ (\ref{29}), ${\cal Y}$ (\ref{30}), ${\cal
M}$ (\ref{31}) (see Fig. 3), and the data given by tables 2, 3, we
infer that the total
 yield of activity by the end of exposition would be

\begin{equation} {\cal Y}(T_e,J_e,{\cal A}_{bn})=
J_e \cdot T_e \cdot {\cal A}_{nb} \cdot 1.7 \cdot 10^{11}
\mbox{kBq}\approx 5 \cdot 10^3 \cdot J_e \cdot T_e \cdot
{\cal A}_{bn}\, \mbox{Ci}, \label{f1}
\end{equation}
provided the exposition time $T_e\lesssim 15\mbox{h}$. At $T_e=1
\mbox{h}$ and $J_e=1 \mbox{A}{/}{\mbox{cm}^2}$,
 we have got the value (\ref{33}) given above. The accelerator can
be turned on and off at will and without any consequences, and
exchanging the irradiated targets is rather a simple thing, which
are apparent advantages of the accelerator-based radio-isotope
production over the routine reactor-based one. So we are in
position of irradiating a set of
 $\mbox{Mo}$ targets successively, one after the other with the
exposition time of each one equal to the most efficient value
$T_e\approx 15 \mbox{h}$. Then, the yield of activity produced in
every one of those samples is

\begin{equation}
{\cal Y}(15\mbox{h},J_e,{\cal A}_{bn})\approx 5 \cdot 10^3 \cdot
J_e
\cdot 15
\cdot {\cal A}_{nb}\, \mbox{Ci}, \label{f2}
\end{equation}
As understood, this way leads to the greatest yield of activity
generally attainable during a total given exposition time. Let we
irradiate in this manner a set of ten samples so that the total
exposition time consumed in one week constitutes $\approx 150
\mbox{h}$. That series of ten targets expositions results in the total
yield of activity per one week

\begin{equation}
{\cal Y}(10{\times}15\mbox{h},J_e,{\cal A}_{bn})\approx 5 \cdot
10^3
\cdot J_e
\cdot 15{\times}10
\cdot {\cal A}_{nb}\, \mbox{Ci}, \label{f3}
\end{equation}
As observed, this total activity is accumulated in the separated
samples, taken together. On the other hand, we can continually
irradiate one single sample one week long, {\it i.e.} again for
$\approx150\mbox{h}$. In this case the evaluation accordingly
Sections 3, 4 (Fig. 3, table 3) results in the total yield of
activity accumulated per one week in that single target

\begin{equation}
{\cal Y}(150\mbox{h},J_e,{\cal A}_{bn})\approx 3.2\cdot 10^3 \cdot
J_e
\cdot 150\cdot {\cal A}_{nb}\, \mbox{Ci}, \label{f4}
\end{equation}
which shows up to be noticeably smaller than the quantity
(\ref{f3}). Let us opt a realizeable value
$J_e=10\mbox{mA}{/}{\mbox{cm}^2}$, and presume that we operate
targets composed of the pure isotope $^{100}\mbox{Mo}$, {\it i.e.}
${\cal A}_{bn}=1$ in the expressions (\ref{f1})-(\ref{f4}). Then we
arrive at

\begin{equation}
{\cal Y}(10{\times}15\mbox{h},10\mbox{mA},1)\approx 7.5
\cdot 10^3 \mbox{Ci},
\label{f5}
\end{equation}

\begin{equation}
{\cal Y}(150\mbox{h},10\mbox{mA},1)\approx 4.8\cdot10^3\mbox{Ci}.
\label{f6}
\end{equation}
Recalling $\tau_{^{99}\mbox{Mo}}=96\mbox{h}$, the aforesaid {\it
6-day curies} activities corresponding to the quantities
(\ref{f5}), (\ref{f6}) are directly evaluated

\begin{equation}
{\cal Y}_{\it 6-day}(10{\times}15\mbox{h},10\mbox{mA},1)\approx
1.67
\cdot 10^3 \mbox{Ci},
\label{f7}
\end{equation}

\begin{equation}
{\cal Y}_{\it 6-day}(150\mbox{h},10\mbox{mA},1)\approx
1.07\cdot10^3\mbox{Ci}.
\label{f8}
\end{equation}
These quantities (\ref{f5})-(\ref{f8}), evaluated with practicable
$^{100}\mbox{Mo}$ targets and a realizable electron beam, prove to
be competitive with the marketable {\it large-scale} productivity
of the {\it large-scale producers}, who furnished on the market
more than $1000$ {\it 6-day curies} of $^{99}\mbox{Mo}$ {\it per
week}, recovered on the routine reactor basis with operating $HEU$
targets \cite{hu}.

Nowadays, the pure laboratory study \cite{m1,m2,m3,mas3} of the
processes (\ref{32}), (\ref{34}), (\ref{36}), with using foils of
natural $\mbox{Mo , Sn , U}$ as irradiated targets, have blazed the
trail towards {\it the large-scale} radio-isotope manufacturing
based on the electron-accelerator driving photo-neutron nuclear
reactions.

In reactor-based radio-isotope producing, no matter whether the
$HEU$- or $LEU$-target is used, the quantity of $^{99}\mbox{Mo}$
available for sale and harnessing is much less than the total
quantity of $^{99}\mbox{Mo}$ produced in an irradiated target
because of this 6 day delay and, primarily, because of losses
caused by the very sophisticated and time-consuming target
processing, still before shipment to the market. Upon irradiating
and then cooling, the targets are processed into {\it hot cells
facilities}, which can cost as much as tens of millions of dollars
to construct, and which are very sophisticated to operate
\cite{int8,hu}. The point is to recover a thoroughly purified
desired radio-isotope ( for instance
 $^{99}\mbox{Mo}$) out of a target where this constitutes, at best,
a few per cent among $^{235}\mbox{U}$ fission fragments \cite{fis}.
In particular, the isotope $^{99}\mbox{Mo}$, several hours after a
moment of fission, constitutes $\sim 6\%$ of fission fragments. The
other way round, in the accelerator-based photo-neutron production,
there requires no {\it hot-sells} and subsidiary equipment for
targets processing after the end of exposition, as a matter of
fact. As well, the desired isotope, {\it e.g.} $^{99}\mbox{Mo}$, is
immediately elaborated within a pure molibdenium target, so that
there is no need to purify anything and manage any wastes.
Therefore the aforesaid 6-day term for calibrating activity of the
shipment ({\it 6-day curies}) is to be recounted just from the end
of target exposition.

Used either $HEU$- or $LEU$-target, one of the most important issue
to take care of is anyway to eliminate, or at least to minimize,
the weapon-usable waste streams resulting from radio-isotope
production. In particular concern is that little or no progress is
being made for now in this way. Only a very small fraction,
typically about $3\%$, of the $^{235}\mbox{U}$ in a target
undergoes fission in reactor-based radio-isotope, {\it e.g.}
$^{99}\mbox{Mo}$, producing \cite{int8,hu}. The vast majority of
the uranium in the target, along with other fission products and
target materials, are eventually treated as wastes, no matter
whether the $HEU$- or $LEU$-target is used. Tens of kilograms of
$HEU$ wastes are annually accumulated worldwide. By contrast, the
electron-accelerator-based radio-isotope production has nothing to
do with any radioactive wastes, there is no waste stream at all, in
actual fact.

The decay product of $^{99}\mbox{Mo}$, the isotope
$^{99m}\mbox{Tc}$ (see Section 5), we are primarily focused on, is
used in about two-thirds of all the diagnostic and therapeutic
nuclear-medical procedures all over the world
\cite{ind,int2,int8,nru}. The metastable radio-isotope
$^{99m}\mbox{Tc}$ having got the short life-time, the
$^{99}\mbox{Mo}$ recovered out of an irradiated target is shipped
to radio-pharmacies and hospitals within the {\it technetium
generators} that are {\it eluted} to obtain the desired
$^{99m}\mbox{Tc}$ at destinations. The calibration to asses {\it
Tc-generator} activity is based on the number of curies that are
contained in a {\it generator} on the day of, or the day after its
delivery to an user \cite{int7,int8,hu}. In the considered
photo-neutron radio-isotope production, an irradiated $\mbox{Mo}$
target could be processed in the {\it Tc-generator} just after the
end of exposition, which is again an evident advantage over the
routine reactor-based production, as the last requires a
considerable time and work to prepare that irradiated $HEU$- or
$LEU$-target for usage in the {\it Tc-generator}. Of course, the
appropriate required {\it Tc-generator} is anew to be designed and
built for recovering $^{99m}\mbox{Tc}$ from an irradiated
$\mbox{Mo}$ sample, upon photo-producing the $^{99}\mbox{Mo}$
radio-isotope therein, instead of the routine {\it Tc-generator} to
process the blend of different $\mbox{Mo}$ isotopes extracted out
of $^{235}\mbox{U}$ fission fragments. The findings explicated in
Section 5 will serve for all these actual purposes. The method how
to recover the $\mbox{Tc}$ from $\mbox{Mo}$ irradiated targets are
by now elaborated as well \cite{is1,pat}.

 As readily understood \cite{hu}, refurbishing the obsolent
reactors and converting from $HEU$-based to $LEU$-based
radio-isotope production would be anyway time consuming, at least
5-6 years long, and technically sophisticated, at least not less
than to bring online the accelerator-based radio-isotope
production. Along with other complexities, the conversion would
anew require the very expensive construction of the special {\it
hot-cells} for processing now the $LEU$-targets, which are not
involved at all into the accelerator-based photo-neutron
radio-isotope production.

In the routine radio-isotope reactor-based elaboration, there is a
very long implicated and vulnerable supply chain of manifold
operations, beginning with the $HEU$ provision of targets producers
and terminating in treating end-users. Any contingency in a single
link results in a fatal malfunction of all the chain, all the more
that diverse operations are performed at different sites, in
different countries, and even at different continents. In contrast,
radio-isotope photo-neutron production would be accomplished at one
single site: from an unsophisticated target preparation
straightforwardly to radio-pharmaceutical manufactures, or, as {\it
e.g.} in the $^{99}\mbox{Mo}$ production case, to charging the {\it
Tc-generator}. There is no issue of shipment of the
 $^{99}\mbox{Mo}$ product to the $^{99m}\mbox{Tc}$ {\it generator}
manufacturing facilities. The losses of radio-isotope yield caused
by decay rate would be then minimized, and even almost eliminated,
by co-locating all the engaged facilities. Under such
circumstances, any irradiated $^{100}\mbox{Mo}$ target, upon
utilizing by {\it Tc-generator}, would be restored, and then
exposed anew. A circle of this kind could many times be repeated
which would allow saving the stick of starting enriched material,
as {\it e.g.} the $^{100}\mbox{Mo}$ isotope in producing the
$^{99}\mbox{Mo}$ isotope. That agenda would offer the possibility
of self-contained generator systems being feasible for central
radio-pharmaceutical labs for a grope of hospitals. So, for all we
have acquired, there offers a new stream from $^{99}\mbox{Mo}$
production to an end-user consumption of {\it kits} prepared with
$^{99m}\mbox{Tc}$.

As understood, see Section 4, the nuclear photo-neutron process
$A(Z,N)(\gamma,n)A'(z,N-1)$ can serve to produce the great variety
of desired radio-isotopes, not only with $A\sim 100$, $A\sim 140$,
{\it i.e.} not only with $A$ placed within the humps of mass
distribution of uranium fission fragments. Especially, it is to
indicate the isotope $^{201}\mbox{Tl}$, that is thought to replace
$^{99}\mbox{Mo}{/}^{99m}\mbox{Tc}$ usage, and a number of specific
radio-isotopes used in the positron emission tomography $(PET)$,
$^{13}\mbox{N}, \, ^{18}\mbox{F}, \,
^{82}\mbox{Sr}{/}^{82}\mbox{Rb}, \,
^{45}\mbox{Ti}, \, ^{60}\mbox{Cu}$, and so on \cite{pet},
which are not available in the
uranium fission process.

The above discussions are leading to consideration of building new
and operating to-day existing electron linear accelerators
(e-linacs) for the radio-isotope production. As understood, see
Eqs. (\ref{f1})-(\ref{f8}), the for now deployed $1{/}2MW$ e-linac
$TRIUMF$ \cite{int8,tri}, with electron energy $E_e=50 \,
\mbox{MeV}$ and current density $J_e=10\mbox{mA}{/}{\mbox{cm}^2}$
can be considered to be eligible for manufacturing $1000$ {\it
6-day curies} $^{99}\mbox{Mo}$ {\it activity per week}, {\it i.e.}
for the {\it large scale} $^{99}\mbox{Mo}$ production. So, a single
machine of this kind could manufacture and supply the
$^{99}\mbox{Mo}$ marketable quantity comparable to productivity of
a single {\it large scale} producer, provided the required
$^{100}\mbox{Mo}$ pure targets are available. The manifold
efficient methods and technologies to separate various isotopes are
by now developed and deployed, see {\it e.g.} Refs. \cite{is1,isi}.
So, recovery of a pure singe-isotope target, {\it e.g.}
 the pure $^{100}\mbox{Mo}$ isotope target, is believed to be not
over-expensive, when the {\it large scale} radio-isotope
photo-neutron manufacture would be validated.

Laboratories around the world, such as $TRIUMPF$ (Canada), $BNL$,
$ORNL$ (U.S.), $IRN$-Orseay, $GANIL$ (France), $ELBE$ (Deutshland)
 \cite{tri,onrl,elb}, have expertise and facilities that can be
used immediately. The construction of far more powerful e-linacs,
with $E_e=50 \, \mbox{MeV}, J_e=100 \mbox{mA}{/}{\mbox{cm}^2}$, are
by now under way, and they are believed to be commissioned before
long, in 3-4 years, which would underpin a long-term enhancement of
desired radio-isotopes production \cite{int8}. The estimated cost
of such new accelerator is about \$50 millions, whereas the cost of
an research and test reactor, with the thermal neutron flux
$\approx (10^{14}-10^{15})\mbox{n}{/}{(\mbox{s cm}^2)}$, is
assessed to be at least \$300-\$400 millions. For instance,
construction cost for the new $100\mbox{MW}$ reactor in Cadarache
(France) is estimated to be about 500 millions of euros \cite{hu}.
Just fixing the $MAPLE$ reactors would have cost tens of millions
of dollars, whereas building such new reactor or refurbishing $NRU$
would have cost hundreds of millions of dollars. Even 5-years
license extension for $NRU$ would have required expenditure of
about hundred of millions of dollars \cite{int8,hu}. Similarly, the
reactor operations are far more costly than the e-linac operations.
Power consumption, dominating the operating costs, is roughly
estimated to be a few $MW$s for an aforesaid e-linac, and the total
operating costs are assessed to constitute about $10\%$ of the
capital investment. Such an accelerator facility, as treated above,
would be viewed as a single-purposed facility operating strictly
for radio-isotope production business.

The $\gamma$-flux, converted from electron beam of an electron
accelerator, can also be used for the photo-fission of
$^{238}\mbox{U}$ , utilizing the natural or depleted uranium
targets, with subsequent recovering the desired radio-isotope, {\it
e.g.} $^{99}\mbox{Mo}$ , from the fission fragments blend
\cite{int8}. The heretofore treated photo-neutron production of
radio-isotopes is certain to be far more efficient and viable than
the radio-isotope recovery out of the fission fragments of the
$^{238}\mbox{U}$ photo-fission. In order to grasp the reason one is
first to recollect that cross-sections $\sigma_{\gamma n}$ of the
reaction (\ref{00}) are akin the cross-section $\sigma_{\gamma f}$
of the $^{238}\mbox{U}$ photo-fission, both being of the same order
\cite{238}. Target nuclear densities are of the same order as well.
Yet the desired radio-isotope, {\it e.g.} $^{99}\mbox{Mo}$ ,
constitutes at best only a few per cent, never more than $\sim
(5-6)\%$, among $^{238}\mbox{U}$ photo-fission fragments. Thus,
given the $\gamma$-flux causing both reactions is the same, the
yield of photo-neutron reaction proves to be far more abundant than
one from processing the irradiated $^{238}\mbox{U}$ sample upon the
$^{238}\mbox{U}$ photo-fission. Anyway, the anew target processing
and generator-manufacturing facilities are to be deployed, in using
both photo-production, $^{100}\mbox{Mo}(\gamma , n)^{99}\mbox{Mo}$,
and photo-fission $^{238}\mbox{U}$ process,
$^{238}\mbox{U}(\gamma,\mbox{F})$. Yet although by this
$^{99}\mbox{Mo}$ photo-fission production, just alike in the
photo-neutron process, the accelerator does not produce radioactive
waste from its operation, the waste from the irradiated targets
chemical processing to recover, extract and purify $^{99}\mbox{Mo}$
would be similar to the waste of the reactor-based production. The
$^{99}\mbox{Mo}$ recovery from $^{238}\mbox{U}$ photo-fission
fragments and from $^{235}\mbox{U}$ thermal-neutron fission
fragments are actually analogous. In either methods, the capital
investments emerge to be very high, as the special facilities, in
particular the {\it hot-sells}, are required to deal with the
emission and disposal of highly radioactive fission products. In
contrast, in the radio-isotope photo-neutron production, there are
neither uranium fission fragments to deal with, nor radioactive
wastes at all. Treated photo-neutron radio-isotope producing, there
occur no secondary trance-uranic nuclei, in particular
$^{239}\mbox{Pu}$, associated with radio-isotope producing based on
the $^{238}\mbox{U}$ photo-fission. As understood, see Eqs.
(\ref{f1})-(\ref{f8}), the mass of about $20\mbox{g}$ of $100\%$
enriched $^{100}\mbox{Mo}$ is quite sufficient to design a
practicable target for the large-scale $^{99}\mbox{Mo}$
photo-neutron production, whereas an depleted $^{238}\mbox{U}$
target for the photo-fission radio-isotope production would require
the mass of at least $\sim 200\mbox{g}$ \cite{int8}. This large
target mass is a substantial challenge in procuring required purity
of the obtained medical radio-isotopes. Contamination possibility
of a produced radio-isotope, {\it e.g.} $^{99}\mbox{Mo}$ , even
increases due to the much larger mass of the photo-fission
$^{238}\mbox{U}$ target as compared to the mass of the
$^{235}\mbox{U}$ target in the case of the thermal neutron fission.
So, as it has been coming to light, the photo-neutron production
techniques has got a lot of advantages over the photo-fission one.
Surely, it is least of all to say that these two approaches rule
each other out. They are naturally to complement each other to
procure the most reliable radio-isotope supply.

Further calculations and laboratory measurements are required to
verify and validate the proof-of-principles that have been treated
in the work presented.
\section*{ A\lowercase{cknowledgments}}
Authors are thankful to O.D. Maslov, Z. Panteleev, Yu. P. Gangrsky
for the valuable discussions.

\end{linenumbers}
\newpage
\begin{figure}[p]
\begin{center}
\begin{picture}(120,125)(0,0)
\SetColor{Red}
\SetWidth{2.0} \LongArrow(0,95)(0,55)
\Line(120,95)(120,55)
\Line(0,95)(120,95)
\Line(0,55)(120,55)
\SetColor{Blue}
\Line(0,35)(120,35)
\Line(0,-5)(120,-5)
\Line(120,35)(120,-5)
\LongArrow(0,35)(0,-5)
\SetColor{ForestGreen}
\ArrowLine(60,130)(60,96)
\SetColor{OliveGreen}
\LongArrow(60,96)(60,76)
\SetColor{Plum}
\Photon(60,76)(60,15){2}{10}
\CCirc(60,12){3}{Plum}{Orange}
\Text(-10,95)[]{${\textRed\bf 0}$}
\Text(-13,55)[]{${\textRed\bf R_C}$}
\Text(-10,75)[]{$\bf x$}
\Text(30,65)[]{$\mbox{W}$}
\rText(128,82)[][l]{$\bf converter$}
\Text(-10,35)[]{${\textBlue\bf 0}$}
\Text(-13,-3)[]{${\textBlue\bf R_S}$}
\Text(-10,15)[]{$\bf y$}
\Text(29,2.5)[]{$\mbox{Mo,Sn,U}$}
\rText(128,15)[][l]{$\bf sample$}
\Text(35,110)[]{${\textPineGreen\bm{\rho_e(E_e)}}$}
\Text(85,110)[]{${\textPineGreen\bm{J_e(t)}}$}
\Text(35,85)[]{${\textOliveGreen\bm{E_e(x)}}$}
\Text(88,65)[]{${\textPlum\bm{J_\gamma(x,k)}}$}
\Text(88,25)[]{${\textPlum\bm{J_\gamma(y,k)}}$}
\Text(55,-0)[]{${\textBlack .}$}
\end{picture}\\
\end{center}
\begin{center}
\*{{\bf Fig. 1.} The setup scheme.}
\end{center}
\label{figure1}
\vspace{16cm}
\end{figure}

\begin{figure}[p]
\begin{center}
\begin{picture}(440,445)(0,0)
\LinAxis(0,0)(420,0)(2,10,5,0,1.5)
\Text(0,-9)[]{\large$\bf 8$}
\Text(210,-9)[]{\large$\bf 18$}
\Text(420,-9)[]{\large$\bf 28$}
\Text(210,-27)[]{\large$\bf k=E_{\bfgr\gamma}(\mbox{MeV})$}
\LinAxis(0,435.7)(420,435.7)(2,10,-5,0,1.5)
\SetColor{Red}
\LinAxis(0,0)(0,435.7)(7.643,4,-5,0,1.5)
\Text(-11,0)[]{\textRed \large$\bf 0$}
\Text(-11,57)[]{\large$\bf 1$}
\Text(-11,114.012)[]{\large$\bf 2$}
\Text(-11,171.02)[]{\large$\bf 3$}
\Text(-11,228.03)[]{\large$\bf 4$}
\Text(-11,288.03)[]{\large$\bf 5$}
\Text(-11,342.04)[]{\large$\bf 6$}
\Text(-11,399.04)[]{\large$\bf 7$}
\rText(-30,225)[][l]{\Large${\bf J_{\bfgr\gamma C}}
{\times}{10^{17}}{/}{\mbox{cm}^2\mbox{s}\mbox{MeV}}$}
\Text(382,100)[]{$\bf \bar E_e=100$}
\Text(352,80)[]{$\bf R_{\mbox{W}}=0.4$}
\Text(25,315)[]{$\bf {\bar E_e}{=}{50}$}
\Text(37,300)[]{$\bf {R_{\mbox{W}}}{=}{0.3}$}
\Text(27,160)[]{$\bf {\bar E_e}{=25}$}
\Text(40,146)[]{$\bf {R_{\mbox{W}}}{=}{0.3}$}
\Text(83,30)[]{$\bf {\bar E_e}{=20}$}
\Text(113,15)[]{$\bf {R_{\mbox{W}}}{=}{0.1}$}
\SetColor{Blue}
\LinAxis(420,0)(420,435.7)(4,10,5,0,1.5)
\Text(433,108.925)[]{{\textBlue \large$\bf 50$}}
\Text(430,0)[]{{\textBlue \large$\bf 0$}}
\Text(433,217.85)[]{{\textBlue \large$\bf 100$}}
\Text(433,326.775)[]{{\textBlue \large$\bf 150$}}
\Text(433,435.7)[]{{\textBlue \large$\bf 200$}}
\rText(440,157)[][l]{{\textBlue
\Large$\bm{\sigma_{\gamma n}(\mbox{mb})}$}{\textBlack .}}
\SetScale{2.2}
\SetWidth{0.8}
\SetScaledOffset(-0.7,0.02)
\Curve{(4.660,11.2500)(7.380,12.3500)(10.100,17.6500)
(12.820, 22.9000)(15.540,23.9000)(18.260,23.7500)(20.980,
24.5500)(23.700,30.8000)(26.420,31.2500)(29.140,31.4500)
(31.860,40.4500)(34.580,47.7500)(37.300,55.6500)(40.020
,67.4500)(42.740,79.3000)(45.460,87.8500)(48.180,
97.2000)(50.900,107.9500)(53.620,119.3000)(56.340,126.1500)
(59.060,134.2500)(61.780,152.4500)(64.500,157.8500)(67.220,
149.8000)(69.940,138.8000)(72.660,125.9000)(75.380,112.2500)
(78.100,100.2000)(80.820,89.3000)(83.540,74.1000)
(86.260,60.3500)(88.980,57.5500)(91.700,57.5000)
(94.420,54.2000)(97.140,47.6500)(99.860,
30.0500)(102.580,25.3000)(105.300,29.2000)(108.020,
21.9500)(110.740,12.8000)(113.460,13.7500)(116.180,17.4000)
(118.900,12.0000)(121.620,5.6500)(124.340,8.2500)(127.060,11.6000)
(129.780,12.4500)(132.500,11.0500)(135.220,11.2500)(137.940,
9.3000)(140.660,-0.3500)(143.380,2.0500)(146.100,11.9000)(148.820,
12.6000)(151.540,9.6000)(154.260,3.3000)(156.980,6.3500) (159.700,
11.5500)(162.420,10.1000)(165.140,7.4500)(167.860,5.3500)(170.580,
13.9500)(173.300,13.4000)(176.020,8.8000)(178.740,18.4000)
(181.460,23.1000)(184.180,17.6000)(186.900,14.4500)}
\SetColor{Red}
\DashCurve{(4.660,64.87988)(7.380,61.35512)
(10.100,57.96339)
(12.820,53.49026)(15.540,50.73612)(18.260,48.06006)
(20.980,44.51120)(23.700,42.29418)(26.420,40.11268)
(29.140,37.36500)
(31.860,35.50480)(34.580,33.21210)(37.300,31.22510)(40.020,
29.40667)(42.740,27.64565)(45.460,26.33207)(48.180,
24.58541)(50.900,23.41870)(53.620,21.67001)(56.340,20.67677)
(59.060,19.66472)(61.780,18.04686)(64.500,17.22267)(67.220,
16.37717)(69.940,14.82440)(72.660,14.16278)(75.380,13.48382)
(78.100,12.77408)(80.820,11.43132)(83.540,10.90368)(86.260,
10.35565)(88.980,9.77505)(91.700,8.58883)(94.420,
8.17633)(97.140,7.74387)(99.860,7.28140)(102.580,6.22407)(105.300,
5.90767)(108.020,5.57301)(110.740,5.21205)(113.460,4.28602)
(116.180
,4.04733)(118.900,3.79207)(121.620,3.51353)(124.340,2.74001)
(127.060,2.56237)(129.780,2.36920)(132.500,2.15424)
(135.220,1.56015)
(137.940,1.42900)(140.660,1.28213)(143.380,0.80752)
(146.100,0.72222)
(148.820,0.62432)(151.540,0.31017)(154.260,0.25897)
(156.980,0.07430)
(159.700,0.0000)(162.420,0.0000)(165.140,0.0000)(167.860
,0.0000)(170.580,0.0000)(173.300,0.0000)(176.020,0.0000)(178.740
,0.0000)(181.460,0.0000)(184.180,0.0000)(186.900,0.0000)}{4.0}
\DashCurve{(4.6600,38.6946)(7.3800,36.8275)(10.1000,35.0754)
(12.8200,33.4251)(15.5400,31.8648)(18.2600,30.3832)(20.9800,28.9699)
(23.7000,27.6149)(26.4200,26.3080)(29.1400,25.0392)(31.8600,23.7974)
(34.5800,22.5701)(37.3000,21.3341)(40.0200,19.9124)(42.7400,18.3952)
(45.4600,16.9643)(48.1800,15.6203)(50.9000,14.3597)(53.6200,13.1693)
(56.3400,12.0554)(59.0600,10.9998)(61.7800,9.9845)(64.5000,9.0465)
(67.2200,8.1651)(69.9400,7.3233)(72.6600,6.5343)(75.3800,5.7974)
(78.1000,5.1003)(80.8200,4.4493)(83.5400,3.8373)(86.2600,3.2579)
(88.9800,2.7344)(91.7000,2.2437)(94.4200,1.7922)(97.1400,1.3702)
(99.8600,0.9994)(102.5800,0.6743)(105.3000,0.3835)(108.0200,0.1429)
(110.7400,0.0166)(113.4600,0.0002)(116.1800,0.0000)(118.9000,0.0000)
(121.6200,0.0000)(124.3400,0.0000)(127.0600,0.0000)(129.7800,0.0000)
(132.5000,0.0000)(135.2200,0.0000)(137.9400,0.0000)(140.6600,0.0000)
(143.3800,0.0000)(146.1000,0.0000)(148.8200,0.0000)(151.5400,0.0000)
(154.2600,0.0000)(156.9800,0.0000)(159.7000,0.0000)(162.4200,0.0000)
(165.1400,0.0000)(167.8600,0.0000)(170.5800,0.0000)(173.3000,0.0000)
(176.0200,0.0000)(178.7400,0.0000)(181.4600,0.0000)(184.1800,0.0000)
(186.9000,0.0000)}{4.0}
\DashCurve{(4.6600,135.5630)(7.3800,129.9217)(10.1000,124.6396)
(12.8200,119.6833)(15.5400,115.0226)(18.2600,110.6306)
 (20.9800,106.4829)(23.7000,102.5574)(26.4200,98.8336)(29.1400
,95.2926)(31.8600,91.9166)(34.5800,88.6883)(37.3000,
85.5907)(40.0200,82.6038)(42.7400,79.6595)
(45.4600,76.6932)(48.1800,
73.8043)(50.9000,71.1115)(53.6200,68.3927)(56.3400,66.0611)
(59.0600,63.6359)(61.7800,61.2336)(64.5000,59.2594)
(67.2200,57.3075)
(69.9400,55.0919)(72.6600,53.1779)(75.3800,51.5724)
(78.1000,49.9678)(80.8200,
48.0954)(83.5400,46.3438)(86.2600,44.9905)
(88.9800,43.6964)(91.7000,42.3396)(94.4200,40.7153)
(97.1400,39.3203)(99.8600,38.2354)(102.5800,37.1863)
(105.3000,36.0964)(108.0200,34.7772)(110.7400,33.5190)
(113.4600,32.6016)(116.1800,31.7517)(118.9000,30.8887)
(121.6200,29.9056)(124.3400,28.7365)(127.0600,27.8795)
(129.7800,27.1790)(132.5000,26.4829)(135.2200,25.7379)
(137.9400,24.8266)(140.6600,23.9471)(143.3800,23.3213)
(146.1000,22.7481)(148.8200,22.1518)(151.5400,21.4433)
(154.2600,20.6686)(156.9800,20.0725)(159.7000,19.5859)
(162.4200,19.0855)(165.1400,18.4923)(167.8600,17.8421)
(170.5800,17.3371)(173.3000,16.9102)(176.0200,16.4395)
(178.7400,15.8968)(181.4600,15.4102)(184.1800,15.0131)
(186.9000,14.5767)}{4.0}
\DashCurve{(4.6600,198.3388)(7.3800,190.7345)(10.1000,183.6082
)(12.8200,176.9177)(15.5400,170.6252)(18.2600,164.6973)(20.9800
,159.1042)(23.7000,153.8191)(26.4200,148.8180)(29.1400,144.0791
)(31.8600,139.5829)(34.5800,135.3116)(37.3000,131.2490)(40.0200
,127.3804)(42.7400,123.6924)(45.4600,120.1724)(48.1800,116.8093
)(50.9000,113.5925)(53.6200,110.5124)(56.3400,107.5599)(59.0600
,104.7267)(61.7800,102.0050)(64.5000,99.3874)(67.2200,96.8670)
(69.9400,94.4371)(72.6600,92.0915)(75.3800,89.8240)(78.1000,87.6284)
(80.8200,85.4988)(83.5400,83.4286)(86.2600,81.4104)
(88.9800,79.4250)(91.7000,77.4329)(94.4200,75.4709)
(97.1400,73.5098)(99.8600,71.6330)(102.5800,69.9199)
(105.3000,68.1924)(108.0200,66.3789)(110.7400,64.7517
)(113.4600,63.2953)(116.1800,61.8772)(118.9000,60.3961)(121.6200
,58.7916)(124.3400,57.3277)(127.0600,56.0912)(129.7800,54.9214
)(132.5000,53.7641)(135.2200,52.5877)(137.9400,51.2789)(140.6600
,49.9268)(143.3800,48.7824)(146.1000,47.8134)(148.8200,46.8849
)(151.5400,45.9641)(154.2600,45.0429)(156.9800,44.0908)(159.7000
,43.0131)(162.4200,41.8775)(165.1400,40.9485)(167.8600,40.1815
)(170.5800,39.4562)(173.3000,38.7378)(176.0200,38.0205)(178.7400
,37.2985)(181.4600,36.5477)(184.1800,35.6953)(186.9000,34.7775)}
{4.0}
\Text(210,-0)[]{\textBlack .}
\end{picture}\\
\end{center}

\vspace{1.cm}

\*{{\bf Fig. 2.}
 The dashed curves represent the k-dependence of the $\gamma$-flux
(\ref{16}) at the final edge of converter for various $\bar E_e
(\mbox{MeV})$
 and the most preferable thicknesses $R_W(\mbox{cm})$, which are
plotted alongside the respective curves. The electron energies are
distributed around the given $\bar E_e (\mbox{MeV})$ accordingly
Eq. (\ref{17}) with the $E_e^b , \; E_e^u , \;
\Delta_e$ values chosen as in  table 5.
The initial electron current density
$J_e={1\mbox{A}}{/}{\mbox{cm}^2}$. The solid curve represents
$k$-dependence of the cross-section of reaction in Eq. (\ref{32}).}
\label{fig2}
\end{figure}

\begin{figure}[p]
\begin{center}
\begin{picture}(150,100)(0,0)
\SetOffset(70,0)
\SetWidth{1.5}
\SetScale{1.3}
\LogAxis(-135,0)(170,0)(3.2,5,0,1.3)
\Text(20,-20)[]{$\bf T$}
\Text(-173,-10)[]{$\bf 1h$}
\Text(-50,-10)[]{$\bf 10h$}
\Text(76,-10)[]{$\bf 100h$}
\SetWidth{0.3}
\Line(-135,50)(170,50)
\SetWidth{1.5}
\LinAxis(-135,0)(-135,50)(4,5,-5,0,1.3)
\LinAxis(170,0)(170,50)(4,5,5,0,1.3)
\rText(-188,25)[][l]{$\bf Y$}
\SetOffset(0,-1)
\Curve{(-80.,46.064)(-50,46.060)(-20.2,46.19)(10.3,46.10)
(40,45.01)(70.5,41.47)(100.6,35.6)(130.7,26.92)(160.8,16.97)
(174.8,11.99)(190.,9.06)
(200.6,7.27)(208.5,5.99)(215.2,5.1)(223.,4.3)}
\end{picture}
\end{center}

\hspace{5cm}

\begin{center}
\begin{picture}(150,100)(0,0)
\SetOffset(10,0)
\SetWidth{1.5}
\SetScale{1.3}
\LogAxis(-90,0)(215,0)(3.2,5,0,1.3)
\Text(84,-25)[]{$\bf T$}
\Text(-113,-10)[]{$\bf 1h$}
\Text(10,-10)[]{$\bf 10h$}
\Text(135,-10)[]{$\bf 100h$}
\SetWidth{0.3}
\Line(-90,50)(215,50)
\SetWidth{1.5}
\LinAxis(-90,0)(-90,50)(2,5,-5,0,1.3)
\LinAxis(215,0)(215,50)(2,5,5,0,1.3)
\rText(-130,25)[][l]{${\cal M}$}
\SetOffset(0,-1)
\Curve{(-80.,0.45)(-50.,0.925)(-20.2,1.75)(10.3,3.5)
(40,6.9)(70.5,12.75)(100.6,21.9)(130.7,33.0)(160.8,41.75)
(174.8,44.0)(190.,44.5)
(200.6,44.5)(208.5,44.75)(215.2,44.75)(221.,44.75)}
\end{picture}\\
\end{center}
\vspace{1cm}

\*{{\bf Fig. 3.} Time dependence of the yield of activity
$Y[{\mbox{Bq}}{/}{\bigl(\mbox{h}\cdot
\mu\mbox{A}\cdot\mbox{mg}\,^{100}\mbox{Mo}\bigr)}]$
 and amount ${\cal M}(\mbox{mg} 10^{-1})$ obtained with the
converter thickness $R_W=0.3 \mbox{cm}$ by irradiation of the
natural $^{nat}\mbox{Mo}$ sample, with $1 \mbox{cm}^2$ area and
thickness $R_{\mbox{Mo}}=0.01
\mbox{cm}$ (foil), by electrons with $\bar E_e=25 \, \mbox{MeV} \,
, \; J_e={1 \mbox{A}}{/}{\mbox{cm}}^2$.}
\label{fig3}

\vspace{7cm}

\end{figure}
\newpage

\begin{table}[p]
\*{ \hspace{-14.7cm} {\bf Table 1}\\ The yield of activity
 $Y[{\mbox{kBq}}{/}{\bigl(\mbox{h}\cdot
\mu\mbox{A}\cdot\mbox{mg}\,^{100}\mbox{Mo}\bigr)}]$
and amount ${\cal M}(\mbox{mg} 10^{-2})$ of $^{99}\mbox{Mo}$ at
various converter thickness $R_W(\mbox{cm})$, upon irradiating the
natural $^{nat}\mbox{Mo}$ sample, with $1
\,\mbox{cm}^2$ area and $R_{\mbox{Mo}}=0.01\, \mbox{cm}$
(foil), by electrons with $E_e=25
\, \mbox{MeV} \, , \; J_e={1 \mbox{A}}{/}{\mbox{cm}^2}$, during $1
\mbox{h}$.}

\vspace{0.5cm}

\begin{center}
\begin{picture}(300,75)(0,0)
\SetWidth{1.2}
\Line(0,0)(280,0)
\Line(0,90)(280,90)
\Line(0,0)(0,90)
\Line(280,0)(280,90)
\SetWidth{0.3}
\Line(0,30)(280,30)
\Line(0,60)(280,60)
\Text(60,45)[]{3.63}
\Text(20,45)[]{$Y$}
\Text(100,45)[]{4.09}
\Text(140,45)[]{4.07}
\Text(180,45)[]{3.99}
\Text(220,45)[]{3.80}
 \Text(260,45)[]{3.62}
\Text(20,75)[]{$R_{\mbox{W}}$}
\Text(60,75)[]{0.1}
\Text(100,75)[]{0.15}
\Text(140,75)[]{0.175}
\Text(180,75)[]{0.2}
\Text(220,75)[]{0.25}
\Text(260,75)[]{0.30}
\Text(20,15)[]{${\cal M}$}
\Text(60,15)[]{0.18}
\Text(100,15)[]{0.2}
\Text(140,15)[]{0.2}
\Text(180,15)[]{0.196}
\Text(220,15)[]{0.187}
\Text(260,15)[]{0.177}
\end{picture}
\end{center}
\end{table}
\begin{table}[p]
\*{ \hspace{-15.2cm} {\bf Table 2}\\
The yield of activity $Y[{\mbox{kBq}}{/}{\bigl(\mbox{h}\cdot
\mu\mbox{A}\cdot\mbox{mg}\,^{100}\mbox{Mo}\bigr)}]$
 and amount ${\cal M}(\mbox{mg} 10^{-2})$ of the $^{99}\mbox{Mo}$
at various sample thickness $R_{\mbox{Mo}}(\mbox{cm})$, upon
irradiating the natural $^{nat}\mbox{Mo}$ sample during $1\mbox{h}$
by electrons with $E_e=25 \, \mbox{MeV} \, , \; \;
J_e={1\mbox{A}}{/}{\mbox{cm}^2}$, with the converter thickness
$R_W=0.15 \, \mbox{cm}$.}
\label{table2}
\vspace{-0.8cm}
\begin{center}
\begin{picture}(300,100)(0,0)
\SetWidth{1.2}
\Line(0,0)(280,0)
\Line(0,90)(280,90)
\Line(0,0)(0,90)
\Line(280,0)(280,90)
\SetWidth{0.3}
\Line(0,30)(280,30)
\Line(0,60)(280,60)
\Text(60,45)[]{4.09}
\Text(20,45)[]{$Y$}
\Text(100,45)[]{3.75}
\Text(140,45)[]{3.44}
\Text(180,45)[]{3.17}
\Text(220,45)[]{2.93}
\Text(260,45)[]{2.186}
\Text(20,75)[]{$R_{\mbox{Mo}}$}
\Text(60,75)[]{0.01}
\Text(100,75)[]{0.5}
\Text(140,75)[]{1.0}
\Text(180,75)[]{1.5}
\Text(220,75)[]{2.0}
\Text(260,75)[]{4.0}
\Text(20,15)[]{${\cal M}$}
\Text(60,15)[]{0.2}
\Text(100,15)[]{9.2}
\Text(140,15)[]{16.8}
\Text(180,15)[]{23.4}
\Text(220,15)[]{28.65}
\Text(260,15)[]{42.9}
\end{picture}
\begin{picture}(223,100)(0,0)
\rText(-20,45)[][l]{\bf continuing}
\SetWidth{1.2}
\Line(0,0)(239,0)
\Line(0,90)(239,90)
\Line(0,0)(0,90)
\Line(239,0)(239,90)
\SetWidth{0.3}
\Line(0,30)(239,30)
\Line(0,60)(239,60)
\Text(60,45)[]{1.69}
\Text(20,45)[]{$Y$}
\Text(100,45)[]{1.37}
\Text(140,45)[]{1.12}
\Text(180,45)[]{0.948}
\Text(220,45)[]{0.818}
\Text(20,75)[]{$R_{\mbox{Mo}}$}
\Text(60,75)[]{6.0}
\Text(100,75)[]{8.0}
\Text(140,75)[]{10.0}
\Text(180,75)[]{12.0}
\Text(220,75)[]{14.0}
\Text(20,15)[]{${\cal M}$}
\Text(60,15)[]{49.73}
\Text(100,15)[]{53.3}
\Text(140,15)[]{55.0}
\Text(180,15)[]{56.2}
\Text(220,15)[]{56.8}
\end{picture}
\end{center}
\end{table}
\newpage
\begin{table}[p]
\*{ \hspace{-15.5cm} {\bf Table 3}\\
The same as in table 2, yet for $R_W=0.3\, \mbox{cm}$, and for the
initial electron energy distributed around $\bar E_e=50\, \mbox
{MeV}$ accordingly Eq. (\ref{17}) with $E_e^b=48.5 \,\mbox{MeV} \,
, \;
\;E_e^u=52.5 \,\mbox{MeV} \, , \; \;\Delta_e=0.5\,
\mbox{MeV}$.}
\label{table3}
\vspace{-1.5cm}
\begin{center}
\begin{picture}(300,115)(0,0)
\SetWidth{1.2}
\Line(0,0)(280,0)
\Line(0,90)(280,90)
\Line(0,0)(0,90)
\Line(280,0)(280,90)
\SetWidth{0.3}
\Line(0,30)(280,30)
\Line(0,60)(280,60)
\Text(60,45)[]{11.73}
\Text(20,45)[]{$Y$}
\Text(100,45)[]{10.73}
\Text(140,45)[]{9.86}
\Text(180,45)[]{9.07}
\Text(220,45)[]{8.37}
\Text(260,45)[]{7.78}
\Text(20,75)[]{$R_{\mbox{Mo}}$}
\Text(60,75)[]{0.01}
\Text(100,75)[]{0.5}
\Text(140,75)[]{1.0}
\Text(180,75)[]{1.5}
\Text(220,75)[]{2.0}
\Text(260,75)[]{2.5}
\Text(20,15)[]{${\cal M}$}
\Text(60,15)[]{0.58}
\Text(100,15)[]{26.3}
\Text(140,15)[]{48.4}
\Text(180,15)[]{66.8}
\Text(220,15)[]{82.2}
\Text(260,15)[]{94.4}
\end{picture}
\end{center}
\end{table}
\newpage
\begin{table}[p]
\*{ \hspace{-15.5cm} {\bf Table 4}\\
The same as in  table 3, yet for $R_W=0.4
\,\mbox{cm}$, and for the initial electron energy distributed
around $\bar E_e=100\, \mbox {MeV}$ accordingly Eq. (\ref{17}) with
$E_e^b=95 \,\mbox{MeV} \, , \; \;E_e^u=105 \,\mbox{MeV} \, , \;
\;\Delta_e=1\, \mbox{MeV}$.}
\label{table4}
\vspace{-1.5cm}
\begin{center}
\begin{picture}(300,115)(0,0)
\SetWidth{1.2}
\Line(0,0)(280,0)
\Line(0,90)(280,90)
\Line(0,0)(0,90)
\Line(280,0)(280,90)
\SetWidth{0.3}
\Line(0,30)(280,30)
\Line(0,60)(280,60)
\Text(60,45)[]{19.45}
\Text(20,45)[]{$Y$}
\Text(100,45)[]{17.80}
\Text(140,45)[]{16.33}
\Text(180,45)[]{15.03}
\Text(220,45)[]{13.84}
\Text(260,45)[]{12.82}
\Text(20,75)[]{$R_{\mbox{Mo}}$}
\Text(60,75)[]{0.01}
\Text(100,75)[]{0.5}
\Text(140,75)[]{1.0}
\Text(180,75)[]{1.5}
\Text(220,75)[]{2.0}
\Text(260,75)[]{2.5}
\Text(20,15)[]{${\cal M}$}
\Text(60,15)[]{0.95}
\Text(100,15)[]{43.7}
\Text(140,15)[]{80.1}
\Text(180,15)[]{110.6}
\Text(220,15)[]{136.1}
\Text(260,15)[]{157.2}
\end{picture}
\end{center}
\end{table}
\newpage
\begin{table}[p]
\*{ \hspace{-15.8cm} {\bf Table 5}\\
 \hspace{-0.23cm} The yield of activity
 $Y[{\mbox{kBq}}{/}{\bigl(\mbox{h}\cdot
\mu\mbox{A}\cdot\mbox{mg}\,^{100}\mbox{Mo}\bigr)}]$
and amount ${\cal M}(\mbox{mg} 10^{-2})$ of $^{99}\mbox{Mo}$ at
various initial electron energy distributions described by $\bar
E_e \, , \; \; E_e^b \, , \; \; E_e^u \, , \; \; \Delta_e$, all in
$\mbox{MeV}$, as given in Eq. (\ref{17}), and at the most
preferable respective thicknesses $R_W(\mbox{cm})$. The natural
$^{nat}\mbox{Mo}$ sample, with the thickness $R_{\mbox{Mo}}=0.01
\mbox{cm}$ and $1 \mbox{cm}^2$ area, is irradiated by the
electron current $J_e=1 {\mbox{A}}{/}{\mbox{cm}^2}$, during $1
\mbox{h}$.}
\label{table5}
\vspace{-0.6cm}
\begin{center}
\begin{picture}(220,210)(0,0)
\SetWidth{1.2}
\Line(0,0)(199.5,0)
\Line(0,210)(199.5,210)
\Line(0,0)(0,210)
\Line(199.5,0)(199.5,210)
\SetWidth{0.3}
\Line(0,30)(199.5,30)
\Line(0,60)(199.5,60)
\Line(0,90)(199.5,90)
\Line(0,120)(199.5,120)
\Line(0,150)(199.5,150)
\Line(0,180)(199.5,180)
\Text(60,45)[]{2.0}
\Text(20,45)[]{$Y$}
\Text(100,45)[]{4.09}
\Text(140,45)[]{11.73}
\Text(180,45)[]{19.45}
\Text(20,75)[]{$R_{\mbox{W}}$}
\Text(60,75)[]{0.1}
\Text(100,75)[]{0.180}
\Text(140,75)[]{0.3}
\Text(180,75)[]{0.4}
\Text(20,15)[]{${\cal M}$}
\Text(60,15)[]{0.1}
\Text(100,15)[]{0.2}
\Text(140,15)[]{0.58}
\Text(180,15)[]{0.95}
\Text(20,105)[]{$\Delta_e$}
\Text(60,105)[]{0.2}
\Text(100,105)[]{0.2}
\Text(140,105)[]{0.5}
\Text(180,105)[]{1.0}
\Text(20,135)[]{$E_e^u$}
\Text(60,135)[]{20.5}
\Text(100,135)[]{25.5}
\Text(140,135)[]{52.5}
\Text(180,135)[]{105}
\Text(20,165)[]{$E_e^b$}
\Text(60,165)[]{19.5}
\Text(100,165)[]{24.5}
\Text(140,165)[]{48.5}
\Text(180,165)[]{95}
\Text(20,195)[]{$\bar E_e$}
\Text(60,195)[]{20}
\Text(100,195)[]{25}
\Text(140,195)[]{50}
\Text(180,195)[]{100}
\end{picture}
\end{center}
\vspace{1cm}
\end{table}

\newpage
\begin{table}[p]
\*{ \hspace{-15.cm} {\bf Table 6}\\
The same as in table 5, yet here the quantities
 $Y[{\mbox{kBq}}{/}{\bigl(\mbox{h}\cdot
\mu\mbox{A}\cdot\mbox{mg}\,^{118}\mbox{Sn}\bigr)}]$ and ${\cal
M}(\mbox{mg} 10^{-2})$ are obtained for $^{117}\mbox{Sn}^m$
production from the natural tin $^{nat}\mbox{Sn}$ sample, with the
thicknesses $R_{\mbox{Sn}}=0.01
\mbox{cm}$ (foil), $R_{\mbox{Sn}}=2 \mbox{cm}$, and $1
\mbox{cm}^2$ area.}
\label{table6}
\vspace{-1cm}
\begin{center}
\begin{picture}(220,220)(0,0)
\SetWidth{1.2}
\Line(-45,-60)(199.5,-60)
\Line(-45,210)(199.5,210)
\Line(-45,-60)(-45,210)
\Line(199.5,-60)(199.5,210)
\SetWidth{0.3}
\Line(-45,-30)(199.5,-30)
\Line(-45,0)(199.5,0)
\Line(-45,30)(199.5,30)
\Line(-45,60)(199.5,60)
\Line(-45,90)(199.5,90)
\Line(-45,120)(199.5,120)
\Line(-45,150)(199.5,150)
\Line(-45,180)(199.5,180)
\Line(40,-60)(40,210)
\Text(60,45)[]{0.456}
\Text(-3,45)[]{$Y, \, \, R_{\mbox{Sn}}=0.01$}
\Text(100,45)[]{1.04}
\Text(140,45)[]{3.18}
\Text(180,45)[]{5.48}
\Text(-3,75)[]{$R_{\mbox{W}}$}
\Text(60,75)[]{0.1}
\Text(100,75)[]{0.15}
\Text(140,75)[]{0.3}
\Text(180,75)[]{0.4}
\Text(-3,-15)[]{${\cal M}, \, \, R_{\mbox{Sn}}=0.01$}
\Text(60,-15)[]{0.27}
\Text(100,-15)[]{0.62}
\Text(140,-15)[]{1.89}
\Text(180,-15)[]{3.26}
\Text(-3,105)[]{$\Delta_e$}
\Text(60,105)[]{0.2}
\Text(100,105)[]{0.2}
\Text(140,105)[]{0.5}
\Text(180,105)[]{1.0}
\Text(-3,135)[]{$E_e^u$}
\Text(60,135)[]{20.5}
\Text(100,135)[]{25.5}
\Text(140,135)[]{52.5}
\Text(180,135)[]{105}
\Text(-3,165)[]{$E_e^b$}
\Text(60,165)[]{19.5}
\Text(100,165)[]{24.5}
\Text(140,165)[]{48.5}
\Text(180,165)[]{95}
\Text(-3,195)[]{$\bar E_e$}
\Text(60,195)[]{20}
\Text(100,195)[]{25}
\Text(140,195)[]{50}
\Text(180,195)[]{100}
\Text(-3,15)[]{$Y, \, \, R_{\mbox{Sn}}=2.0$}
\Text(60,15)[]{0.34}
\Text(100,15)[]{0.78}
\Text(140,15)[]{2.37}
\Text(180,15)[]{4.08}
\Text(-3,-45)[]{${\cal M}, \, \, R_{\mbox{Sn}}=2.0$}
\Text(60,-45)[]{40.84}
\Text(100,-45)[]{93.00}
\Text(140,-45)[]{283.6}
\Text(180,-45)[]{487.5}
\end{picture}
\end{center}
\vspace{7cm}
\end{table}
\newpage
\vspace{1.8cm}

\begin{table}[p]
\*{ \hspace{-15.6cm} {\bf Table 7}\\
The same as in table 6, yet the quantities
$Y[{\mbox{kBq}}{/}{\bigl(\mbox{h}\cdot
\mu\mbox{A}\cdot\mbox{mg}\,^{238}\mbox{U}\bigr)}]$ and ${\cal
M}(\mbox{mg} 10^{-2})$ are obtained for $^{237}\mbox{U}$ produced
from the natural $^{nat}\mbox{U}$.}
\label{table7}
\vspace{-1cm}
\begin{center}
\begin{picture}(220,225)(0,0)
\SetWidth{1.2}
\Line(-45,-60)(199.5,-60)
\Line(-45,210)(199.5,210)
\Line(-45,-60)(-45,210)
\Line(199.5,-60)(199.5,210)
\SetWidth{0.3}
\Line(-45,-30)(199.5,-30)
\Line(-45,0)(199.5,0)
\Line(-45,30)(199.5,30)
\Line(-45,60)(199.5,60)
\Line(-45,90)(199.5,90)
\Line(-45,120)(199.5,120)
\Line(-45,150)(199.5,150)
\Line(-45,180)(199.5,180)
\Line(40,-60)(40,210)
\Text(60,45)[]{1.01}
\Text(-3,45)[]{$Y, \, \, R_{\mbox{U}}=0.01$}
\Text(100,45)[]{1.77}
\Text(140,45)[]{4.08}
\Text(180,45)[]{6.64}
\Text(-3,75)[]{$R_{\mbox{W}}$}
\Text(60,75)[]{0.15}
\Text(100,75)[]{0.20}
\Text(140,75)[]{0.35}
\Text(180,75)[]{0.45}
\Text(-3,-15)[]{${\cal M}, \, \, R_{\mbox{U}}=0.01$}
\Text(60,-15)[]{6.3}
\Text(100,-15)[]{11.07}
\Text(140,-15)[]{25.44}
\Text(180,-15)[]{41.41}
\Text(-3,105)[]{$\Delta_e$}
\Text(60,105)[]{0.1}
\Text(100,105)[]{0.1}
\Text(140,105)[]{0.5}
\Text(180,105)[]{1.0}
\Text(-3,135)[]{$E_e^u$}
\Text(60,135)[]{20.5}
\Text(100,135)[]{25.5}
\Text(140,135)[]{52.5}
\Text(180,135)[]{105}
\Text(-3,165)[]{$E_e^b$}
\Text(60,165)[]{19.5}
\Text(100,165)[]{24.5}
\Text(140,165)[]{48.5}
\Text(180,165)[]{95}
\Text(-3,195)[]{$\bar E_e$}
\Text(60,195)[]{20}
\Text(100,195)[]{25}
\Text(140,195)[]{50}
\Text(180,195)[]{100}
\Text(-3,15)[]{$Y, \, \, R_{\mbox{U}}=2.0$}
\Text(60,15)[]{0.45}
\Text(100,15)[]{0.78}
\Text(140,15)[]{1.78}
\Text(180,15)[]{2.89}
\Text(-3,-45)[]{${\cal M}, \, \, R_{\mbox{U}}=2.0$}
\Text(60,-45)[]{558.4}
\Text(100,-45)[]{978.9}
\Text(140,-45)[]{2234}
\Text(180,-45)[]{3626}
\end{picture}
\end{center}
\end{table}
\newpage
\begin{table}[p]
\*{ \hspace{-15.2cm} {\bf Table 8}\\ \hspace{-0.565cm}
The time $T_{max}(\mbox
{h})$ of maximum accumulation of the isotope $^{99m}\mbox{Tc}$ and
the respective yield of activity
$Y[{\mbox{kBq}}{/}{\bigl(\mbox{h}\cdot
\mu\mbox{A}\cdot\mbox{mg}\,^{100}\mbox{Mo}\bigr)}]$
and amount ${\cal M}_{\mbox{Tc}^m}(\mbox{mg} 10^{-5})$, as
functions on the time $T_e(\mbox{h})$ of irradiation of the natural
$^{nat}\mbox{Mo}$ sample, with $1 \mbox{cm}^2$ area and
$R_{\mbox{Mo}}=0.01
\mbox{cm}$ (foil), by electrons with $J_e=1
{\mbox{A}}{/}{\mbox{cm}^2} \, , \;
\; \bar E_e=25
\mbox{MeV}$, at the converter thickness $R_W=0.3 \mbox{cm}$.}
\label{table8}
\vspace{-1.3cm}
\begin{center}
\begin{picture}(300,110)(0,0)
\SetWidth{1.2}
\Line(0,-30)(280,-30)
\Line(0,90)(280,90)
\Line(0,-30)(0,90)
\Line(280,-30)(280,90)
\SetWidth{0.3}
\Line(0,0)(280,0)
\Line(0,30)(280,30)
\Line(0,60)(280,60)
\Text(60,45)[]{20}
\Text(20,45)[]{$T_{max}$}
\Text(100,45)[]{23}
\Text(140,45)[]{25}
\Text(180,45)[]{28}
\Text(220,45)[]{30}
\Text(260,45)[]{32}
\Text(20,75)[]{$T_e$}
\Text(60,75)[]{0.5}
\Text(100,75)[]{1}
\Text(140,75)[]{5}
\Text(180,75)[]{10}
\Text(220,75)[]{15}
\Text(260,75)[]{20}
\Text(20,15)[]{$Y_{\mbox{Tc}^m}$}
\Text(60,15)[]{2.54}
\Text(100,15)[]{2.52}
\Text(140,15)[]{2.44}
\Text(180,15)[]{2.32}
\Text(220,15)[]{2.21}
\Text(260,15)[]{2.15}
\Text(20,-15)[]{${\cal M}_{\mbox{Tc}^m}$}
\Text(60,-15)[]{5.65}
\Text(100,-15)[]{11.3}
\Text(140,-15)[]{54.3}
\Text(180,-15)[]{103}
\Text(220,-15)[]{147}
\Text(260,-15)[]{188}
\end{picture}
\end{center}
\end{table}
\vspace{1cm}
\newpage
\begin{table}[p]
\*{ \hspace{-14.cm}
{\bf Table 9}\\ The $T(h)$-dependence of the yield of activity
$Y_{\mbox{Tc}^m}[{\mbox{kBq}}{/}{\bigl(\mbox{h}\cdot
\mu\mbox{A}\cdot\mbox{mg}\,^{100}\mbox{Mo}\bigr)}]$
 and amount ${\cal M}_{\mbox{Tc}_m}(\mbox{mg} 10^{-5})$ of
$^{99m}\mbox{Tc}$ which results inside the natural
$^{nat}\mbox{Mo}$ sample, with $1 \mbox{cm}^2$ area \\
and the
thickness $R_{\mbox{Mo}}=0.01 \mbox{cm}$ (foil), irradiated during
$T_e=0.5
\mbox{h}$ by electrons with $\bar E_e=25 \mbox{MeV}\,, \; J_e=1
{\mbox{A}}{/}{\mbox{cm}^2}$; the
converter thickness $R_W=0.3 \mbox{cm}$.}
\label{table9}
\vspace{-1.3cm}
\begin{center}
\begin{picture}(300,110)(0,0)
\SetWidth{1.2}
\Line(0,0)(280,0)
\Line(0,90)(280,90)
\Line(0,0)(0,90)
\Line(280,0)(280,90)
\SetWidth{0.3}
\Line(0,30)(280,30)
\Line(0,60)(280,60)
\Text(60,45)[]{0.092}
\Text(20,45)[]{$Y_{\mbox{Tc}^m}$}
\Text(100,45)[]{0.434}
\Text(140,45)[]{1.43}
\Text(180,45)[]{2.07}
\Text(220,45)[]{2.54}
\Text(260,45)[]{2.50}
\Text(20,75)[]{$T$}
\Text(60,75)[]{0.5}
\Text(100,75)[]{1.5}
\Text(140,75)[]{5.0}
\Text(180,75)[]{10}
\Text(220,75)[]{20}
\Text(260,75)[]{30}
\Text(20,15)[]{${\cal M}_{\mbox{Tc}^m}$}
\Text(60,15)[]{0.2}
\Text(100,15)[]{0.96}
\Text(140,15)[]{3.19}
\Text(180,15)[]{4.60}
\Text(220,15)[]{5.65}
\Text(260,15)[]{5.56}
\end{picture}
\begin{picture}(223,100)(0,0)
\rText(-20,45)[][l]{\bf continuing}
\SetWidth{1.2}
\Line(0,0)(239,0)
\Line(0,90)(239,90)
\Line(0,0)(0,90)
\Line(239,0)(239,90)
\SetWidth{0.3}
\Line(0,30)(239,30)
\Line(0,60)(239,60)
\Text(60,45)[]{2.32}
\Text(20,45)[]{$Y_{\mbox{Tc}^m}$}
\Text(100,45)[]{2.11}
\Text(140,45)[]{1.9}
\Text(180,45)[]{1.55}
\Text(220,45)[]{1.26}
\Text(20,75)[]{$T$}
\Text(60,75)[]{40}
\Text(100,75)[]{50}
\Text(140,75)[]{60}
\Text(180,75)[]{80}
\Text(220,75)[]{100}
\Text(20,15)[]{${\cal M}_{\mbox{Tc}^m}$}
\Text(60,15)[]{5.16}
\Text(100,15)[]{4.70}
\Text(140,15)[]{4.24}
\Text(180,15)[]{3.44}
\Text(220,15)[]{2.79}
\end{picture}
\end{center}
\vspace{1cm}
\end{table}
\newpage
\begin{table}[p]
\*{ \hspace{-14.cm} {\bf Table 10}\\
The dependence of the yield of $^{99m}\mbox{Tc}$ activity
$Y_{\mbox{Tc}^m}[{\mbox{kBq}}{/}{\bigl(\mbox{h}\cdot
\mu\mbox{A}\cdot\mbox{mg}\,^{100}\mbox{Mo}\bigr)}]$ and
amount ${\cal M}_{\mbox{Tc}^m}(\mbox{mg} 10^{-5})$ on the thickness
$R_{\mbox{Mo}}[\mbox{cm}]$ of the natural $^{nat}\mbox{Mo}$ sample,
with $1 \mbox{cm}^2$ area, irradiated during $T_e=0.5 \mbox{h}$ by
electrons with $J_e=1 {\mbox{A}}{/}{\mbox{cm}^2} \, , \; \; \bar
E_e=25 \mbox{Mev}$. The converter thickness $R_W=0.3 \mbox{cm}$.
All the results are obtained at the time $T=20 \mbox{h}$, counting
from \\ exposition start.}
\label{table10}
\vspace{-1cm}
\begin{center}
\begin{picture}(300,100)(0,0)
\SetWidth{1.2}
\Line(0,0)(280,0)
\Line(0,90)(280,90)
\Line(0,0)(0,90)
\Line(280,0)(280,90)
\SetWidth{0.3}
\Line(0,30)(280,30)
\Line(0,60)(280,60)
\Text(60,45)[]{2.54}
\Text(20,45)[]{$Y_{\mbox{Tc}^m}$}
\Text(100,45)[]{2.30}
\Text(140,45)[]{2.13}
\Text(180,45)[]{1.80}
\Text(220,45)[]{1.36}
\Text(260,45)[]{1.05}
\Text(20,75)[]{$R_{\mbox{Mo}}$}
\Text(60,75)[]{0.01}
\Text(100,75)[]{0.5}
\Text(140,75)[]{1}
\Text(180,75)[]{2}
\Text(220,75)[]{4}
\Text(260,75)[]{6}
\Text(20,15)[]{${\cal M}_{\mbox{Tc}^m}$}
\Text(60,15)[]{5.65}
\Text(100,15)[]{264}
\Text(140,15)[]{478}
\Text(180,15)[]{810}
\Text(220,15)[]{1209}
\Text(260,15)[]{1405}
\end{picture}
\begin{picture}(223,100)(0,0)
\rText(-20,45)[][l]{\bf continuing}
\SetWidth{1.2}
\Line(0,0)(239,0)
\Line(0,90)(239,90)
\Line(0,0)(0,90)
\Line(239,0)(239,90)
\SetWidth{0.3}
\Line(0,30)(239,30)
\Line(0,60)(239,60)
\Text(60,45)[]{0.84}
\Text(20,45)[]{$Y_{\mbox{Tc}^m}$}
\Text(100,45)[]{0.7}
\Text(140,45)[]{0.59}
\Text(180,45)[]{0.45}
\Text(220,45)[]{0.36}
\Text(20,75)[]{$R_{\mbox{Mo}}$}
\Text(60,75)[]{8}
\Text(100,75)[]{10}
\Text(140,75)[]{12}
\Text(180,75)[]{16}
\Text(220,75)[]{20}
\Text(20,15)[]{${\cal M}_{\mbox{Tc}^m}$}
\Text(60,15)[]{1502}
\Text(100,15)[]{1549}
\Text(140,15)[]{1573}
\Text(180,15)[]{1590}
\Text(220,15)[]{1595}
\end{picture}
\end{center}
\end{table}
\newpage
\begin{table}[p]
\*{ \hspace{-14.2cm} {\bf Table 11}\\ The yield of
activity $Y_{\mbox{Tc}^m}[{\mbox{kBq}}{/}{\bigl(\mbox{h}\cdot
\mu\mbox{A}\cdot\mbox{mg}\,^{100}\mbox{Mo}\bigr)}]$
and amount ${\cal M}_{\mbox{Tc}^m}(\mbox{mg} 10^{-5})$ of
$^{99m}\mbox{Tc}$, evaluated at $T=T_e(\mbox{h})$ and
$T=T_{max}(\mbox{h})$, the most preferable time of
$^{99m}\mbox{Tc}$ extraction, for various initial electron energies
$\bar E_e(\mbox{MeV})$. The natural $^{nat}\mbox{Mo}$ sample, with
$1
\mbox{cm}^2$ area and thickness $R_{\mbox{Mo}}=0.01
\mbox{cm}$ (foil), is irradiated during $T_e=1 \mbox{h}$ by the
current $J_e=1 {\mbox{A}}{/}{\mbox{cm}^2}$. \\
$R_W$ is the
converter thickness, preferable at the given $\bar E_e$.}
\label{table11}
\vspace{-1cm}
\begin{center}
\begin{picture}(220,225)(0,0)
\SetWidth{1.2}
\Line(-30,0)(199.5,0)
\Line(-30,210)(199.5,210)
\Line(-30,0)(-30,210)
\Line(199.5,0)(199.5,210)
\SetWidth{0.3}
\Line(-30,30)(199.5,30)
\Line(-30,60)(199.5,60)
\Line(-30,90)(199.5,90)
\Line(-30,120)(199.5,120)
\Line(-30,150)(199.5,150)
\Line(-30,180)(199.5,180)
\Line(40,0)(40,210)
\Text(60,45)[]{0.45}
\Text(5,45)[]{${\cal M}_{\mbox{Tc}^m}(T_e)$}
\Text(100,45)[]{0.91}
\Text(140,45)[]{2.62}
\Text(180,45)[]{4.4}
\Text(5,75)[]{$Y_{\mbox{Tc}^m}(T_{max})$}
\Text(60,75)[]{1.42}
\Text(100,75)[]{2.88}
\Text(140,75)[]{8.28}
\Text(180,75)[]{13.92}
\Text(5,15)[]{${\cal M}_{\mbox{Tc}^m}(T_{max})$}
\Text(60,15)[]{6.29}
\Text(100,15)[]{12.79}
\Text(140,15)[]{36.8}
\Text(180,15)[]{61.82}
\Text(5,105)[]{$Y_{\mbox{Tc}^m}(T_e)$}
\Text(60,105)[]{0.1}
\Text(100,105)[]{0.2}
\Text(140,105)[]{0.59}
\Text(180,105)[]{0.99}
\Text(5,135)[]{$T_{max}$}
\Text(60,135)[]{23}
\Text(100,135)[]{23}
\Text(140,135)[]{23.5}
\Text(180,135)[]{24}
\Text(5,165)[]{$R_{\mbox{W}}$}
\Text(60,165)[]{0.1}
\Text(100,165)[]{0.15}
\Text(140,165)[]{0.3}
\Text(180,165)[]{0.4}
\Text(5,195)[]{$\bar E_e$}
\Text(60,195)[]{20}
\Text(100,195)[]{25}
\Text(140,195)[]{50}
\Text(180,195)[]{100}
\end{picture}
\end{center}
\end{table}
\newpage

\vspace{1cm}

\begin{table}[p]
\*{ \hspace{-8.6cm} \bf Table 12}\\
 \hspace{-2cm} The same as in table 11, yet with $R_{\mbox{Mo}}=2
\mbox{cm}$.
\label{table12}

\vspace{-2.5cm}

\begin{center}
\begin{picture}(220,278)(0,0)
\SetWidth{1.2}
\Line(-30,0)(199.5,0)
\Line(-30,210)(199.5,210)
\Line(-30,0)(-30,210)
\Line(199.5,0)(199.5,210)
\SetWidth{0.3}
\Line(-30,30)(199.5,30)
\Line(-30,60)(199.5,60)
\Line(-30,90)(199.5,90)
\Line(-30,120)(199.5,120)
\Line(-30,150)(199.5,150)
\Line(-30,180)(199.5,180)
\Line(40,0)(40,210)
\Text(60,45)[]{66.4}
\Text(5,45)[]{${\cal M}_{\mbox{Tc}^m}(T_e)$}
\Text(100,45)[]{134}
\Text(140,45)[]{383}
\Text(180,45)[]{643}
\Text(5,75)[]{$Y_{\mbox{Tc}^m}(T_{max})$}
\Text(60,75)[]{1.02}
\Text(100,75)[]{2.06}
\Text(140,75)[]{5.90}
\Text(180,75)[]{9.91}
\Text(5,15)[]{${\cal M}_{\mbox{Tc}^m}(T_{max})$}
\Text(60,15)[]{922}
\Text(100,15)[]{1858}
\Text(140,15)[]{5328}
\Text(180,15)[]{8941}
\Text(5,105)[]{$Y_{\mbox{Tc}^m}(T_e)$}
\Text(60,105)[]{0.07}
\Text(100,105)[]{0.15}
\Text(140,105)[]{0.42}
\Text(180,105)[]{0.71}
\Text(5,135)[]{$T_{max}$}
\Text(60,135)[]{23}
\Text(100,135)[]{23}
\Text(140,135)[]{23.5}
\Text(180,135)[]{24}
\Text(5,165)[]{$R_{\mbox{W}}$}
\Text(60,165)[]{0.1}
\Text(100,165)[]{0.15}
\Text(140,165)[]{0.3}
\Text(180,165)[]{0.4}
\Text(5,195)[]{$\bar E_e$}
\Text(60,195)[]{20}
\Text(100,195)[]{25}
\Text(140,195)[]{50}
\Text(180,195)[]{100}
\end{picture}
\end{center}
\end{table}
\end{document}